\documentclass[12pt]{article}

\newif\ifarxiv
\arxivtrue

\usepackage[margin=1in]{geometry}
\usepackage{setspace}
\ifarxiv
\else
\fi

\usepackage[T1]{fontenc}
\usepackage{lmodern}
\usepackage{amsmath,amssymb}
\usepackage{graphicx}
\usepackage{booktabs}
\usepackage{caption}
\usepackage{subcaption}
\usepackage{float}
\usepackage[numbers,sort&compress]{natbib}
\usepackage{lineno}
\usepackage{xurl}
\usepackage[hidelinks]{hyperref}

\ifarxiv
\else
  \linenumbers
\fi
\newcommand{\papertitle}{Global reanalysis from observations alone with machine learning}

\newcommand{\correspondingauthor}{peter.lean@ecmwf.int}

\newbox{\orcid}\sbox{\orcid}{\includegraphics[scale=0.06]{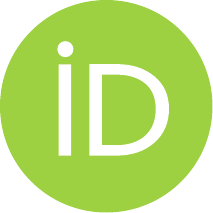}} 

\begin{document}

\begin{center}
  {\LARGE\bfseries \papertitle\par}
  \vspace{0.8em}
  {\href{https://orcid.org/0000-0002-3662-5382}{\usebox{\orcid} Peter Lean}}\textsuperscript{1,*},
  {\href{https://orcid.org/0000-0003-1869-3426}{\usebox{\orcid} Ewan Pinnington}}\textsuperscript{1},
  {\href{https://orcid.org/0000-0003-2808-0463}{\usebox{\orcid} Patrick Laloyaux}}\textsuperscript{1},
  {\href{https://orcid.org/0009-0007-7798-6524}{\usebox{\orcid} Mihai Alexe}}\textsuperscript{2},
  {\href{https://orcid.org/0000-0002-6070-2544}{\usebox{\orcid} Eulalie Boucher}}\textsuperscript{1},
  {\href{https://orcid.org/0000-0003-3952-586X}{\usebox{\orcid} Simon Lang}}\textsuperscript{1},
  Tomas Kral\textsuperscript{1},
  {\href{https://orcid.org/0000-0002-8392-6524}{\usebox{\orcid} Paul Poli}}\textsuperscript{2},
  {\href{https://orcid.org/0000-0001-5330-7071}{\usebox{\orcid} Hans Hersbach}}\textsuperscript{1},
  {\href{https://orcid.org/0000-0001-5302-6093}{\usebox{\orcid} Niels Bormann}}\textsuperscript{1},
  {\href{https://orcid.org/0000-0002-1132-0961}{\usebox{\orcid} Matthew Chantry}}\textsuperscript{1},
  {\href{https://orcid.org/0000-0003-1768-857X}{\usebox{\orcid} Anthony McNally}}\textsuperscript{1}\\[1ex]
 
  \textsuperscript{1}European Centre for Medium-Range Weather Forecasts (ECMWF), Reading, UK\\
  \textsuperscript{2}ECMWF, Bonn, Germany\\  
  \textsuperscript{*}Correspondence: \correspondingauthor
\end{center}

\vspace{1em}

\begin{abstract}

Earth system reanalysis datasets are foundational for weather and climate research and provide the gridded training data used by most machine learning weather prediction systems. Here we show results from a prototype system that suggest that machine learning models trained only on Earth system observations can potentially be used to generate multi-decade global reanalyses without using physics-based numerical models. The resulting gridded fields capture large-scale atmospheric structure and variability across multiple timescales, while exhibiting signs of physical coherence in several key dynamical diagnostics. Evaluations of the prototype against held-out independent atmospheric observations indicate that the root mean square vector error of upper-level winds is close to that of ERA5 when compared at a consistent resolution, and that the standard deviation of the error at the surface is between that of 4th- and 5th-generation ECMWF reanalyses (ERA-Interim and ERA5). Furthermore, while traditional reanalysis production is computationally expensive, typically taking several years to produce, the reanalysis presented here was generated during the course of a single working day. These results suggest that observation-trained machine learning models offer a promising new approach for reanalysis production from observations alone.

\end{abstract}

\section*{Introduction}

Since the reanalysis concept was first introduced in 1988 \citep{TrenberthOlson1988,BengtssonShukla1988}, reanalysis datasets have increasingly served as a cornerstone for weather and climate research and monitoring. Datasets such as ECMWF's ERA5~\citep{hersbach2020era5} (produced in the framework of the EU's Copernicus programme), MERRA-2~\citep{gelaro2017merra2}, JRA-3Q~\citep{Kosaka2024JRA3Q}, and CFSR~\citep{saha2010cfsr} have become de-facto standards, providing trusted information across diverse sectors - from academia to insurance and renewable energy - with substantial economic benefits~\citep{c3s2024era5value}.

These products are generated using data assimilation systems, which statistically combine sparse observational data with prior estimates informed by physical understanding of the Earth system, encoded in numerical models, within a Bayesian framework~\citep{kalnay2002atmospheric,rabier2000ecmwf4dvar}. The underlying numerical model provides the crucial physical constraints needed to fill the gaps between sparse observations, ensuring that the resulting fields are physically consistent across space, time, and geophysical variables.

More recently, these datasets have catalysed the machine learning revolution in weather prediction, with nearly all state-of-the-art data-driven models relying on gridded data assimilation fields for training and initialisation~\citep{keisler2022gnn,pathak2022fourcastnet,bi2023pangu,lam2023graphcast,lang2024aifs,lang2026aifs}. Yet a new frontier of machine learning models trained directly on observational data is rapidly emerging. While hybrid approaches train on a combination of reanalysis and observational data ~\citep{andrychowicz2023metnet3,allen2025endtoend,manshausen2025generative,ni2025huracan,gupta2026healda,pathak2026stormscope}, purely observation-driven techniques such as Direct Observation Prediction (DOP)~\citep{mcnally2024redsky,mcnally2024datadriven,alexe2024graphdop,lean2025learning,boucher2025coupled,laloyaux2025dissect,pinnington2026aifsdop}, and others~\citep{vandal2024earthnet,gong2026eartho1,zhao2026obscast} are trained end-to-end exclusively on observations. DOP models can develop internal latent representations of the Earth system state and dynamics which are independent from the physics encoded in numerical models~\citep{lean2025learning}. Despite the lack of explicit physical priors, such observation-driven models are starting to approach the forecasting skill levels of leading physics-based systems for key surface and upper-air metrics when verified against observations~\citep{pinnington2026aifsdop}.

This raises an interesting scientific question: can observation-driven machine learning models be used to construct physically coherent, gridded reanalysis datasets from sparse observations alone?

Here we address this question by building a purely machine-learning-generated multi-decadal reanalysis dataset constructed using AIFS-DOP \citep{pinnington2026aifsdop}, a variant of ECMWF's Artificial Intelligence Forecasting System (AIFS) \citep{lang2024aifs}. The model is trained end-to-end exclusively from sparse Earth system observations, \textit{with no (re)analysis inputs or targets} and maps sparse observational inputs to predictions of the observed atmospheric state on a regular grid. To produce the reanalysis dataset, we run a series of one-step (six-hour) predictions every six hours, conditioned on recent observations over the previous 30 hours.

We generate 42 years of data at a spatial resolution of approximately 112~km, and show that the resulting dataset captures the large-scale structure of the atmosphere and its variability across multiple timescales. Crucially, we demonstrate that the gridded fields exhibit clear signs of physical coherence and dynamical balance. Finally, evaluation against independent, unassimilated observations indicates that the AI-generated state estimates achieve errors that lie between those of ERA-Interim and ERA5 for some variables, and are comparable to ERA5 for others, highlighting the potential of observation-driven machine learning for reanalysis production from observations alone.

\section*{Model and datasets}

This paper uses the AIFS-DOP model introduced in Pinnington et al. (2026) \citep{pinnington2026aifsdop}. AIFS-DOP builds on the AIFS model, developed and used operationally at ECMWF \citep{lang2024aifs, Moldovan2026}. It draws on experience gained from GraphDOP \citep{alexe2024graphdop}. Sparse observations are represented on a regular O96 input grid, with missing values imputed with zeros after normalisation. The model has an encoder--processor--decoder architecture. The encoder and decoder are graph neural networks (GNNs) with multi-head scaled dot-product attention. After embedding, the input data is mapped by the encoder to the processor grid. The processor consists of 16 pre-norm transformer layers with sliding window attention over longitudinal bands. Finally, the decoder maps the processor-grid embeddings back onto the input grid and into observed variables. The training objective is to predict the observations in the next six-hour window. A mean squared error loss (MSE) which masks grid points with missing values is adopted. The masking prevents the loss from being back-propagated at missing locations to ensure that the model does not learn to predict fill values.

Since the observations in any given six-hour window are incomplete, we adopt a simple cycling procedure inspired by traditional data assimilation. For each initialisation, the model cycles through four six-hour input steps. At each step, the six-hour gridded predictions from the previous step are passed forward and combined with new incoming observations to allow information from earlier windows to inform the state at the current time. Unlike traditional data assimilation, the cycling for each analysis is computed independently to remove serial constraints from the training and inference procedures.

A critical component of a data-driven system is a high quality curated training dataset. Building on ECMWF's experience with curating observational datasets for previous and upcoming reanalysis products, a quality controlled training dataset involving key conventional and satellite instruments spanning 1980-2022 was constructed using the Anemoi framework \citep{lang2024aifs}.

This dataset is the same as that described in Pinnington et al. (2026) \citep{pinnington2026aifsdop}. It consists of satellite and conventional observation sources drawn from ECMWF, EUMETSAT, and NOAA (see Tables~\ref{tab:dataset_description} and \ref{tab:instrument_names}) mapped onto a regular 6-hourly O96 octahedral reduced Gaussian grid \citep{Wedi2014} (approximately 112~km resolution). Grid points which contain no observation information are filled with a missing value which is imputed before entering the model.

The model was trained over the period 1981-2020, with validation between January and May 2021. The reanalysis dataset was generated between January 1st, 1981 and December 31st, 2022. State estimates were made for a set of upper-air variables on standard pressure levels and surface variables every six-hours as detailed in Table \ref{tab:prediction_variables}.

\section*{Atmospheric structure and mean state}

In the following sections, we compare AIFS-DOP state estimates on an O96 grid against ERA5 equivalents at the same resolution. ERA5 is used here as a high quality reference product for assessing large-scale structure, but should not be considered as a ground truth. ERA5 data was not used as a training target in AIFS-DOP.

\begin{figure}[H]
  \centering
  \includegraphics[width=0.75\linewidth]{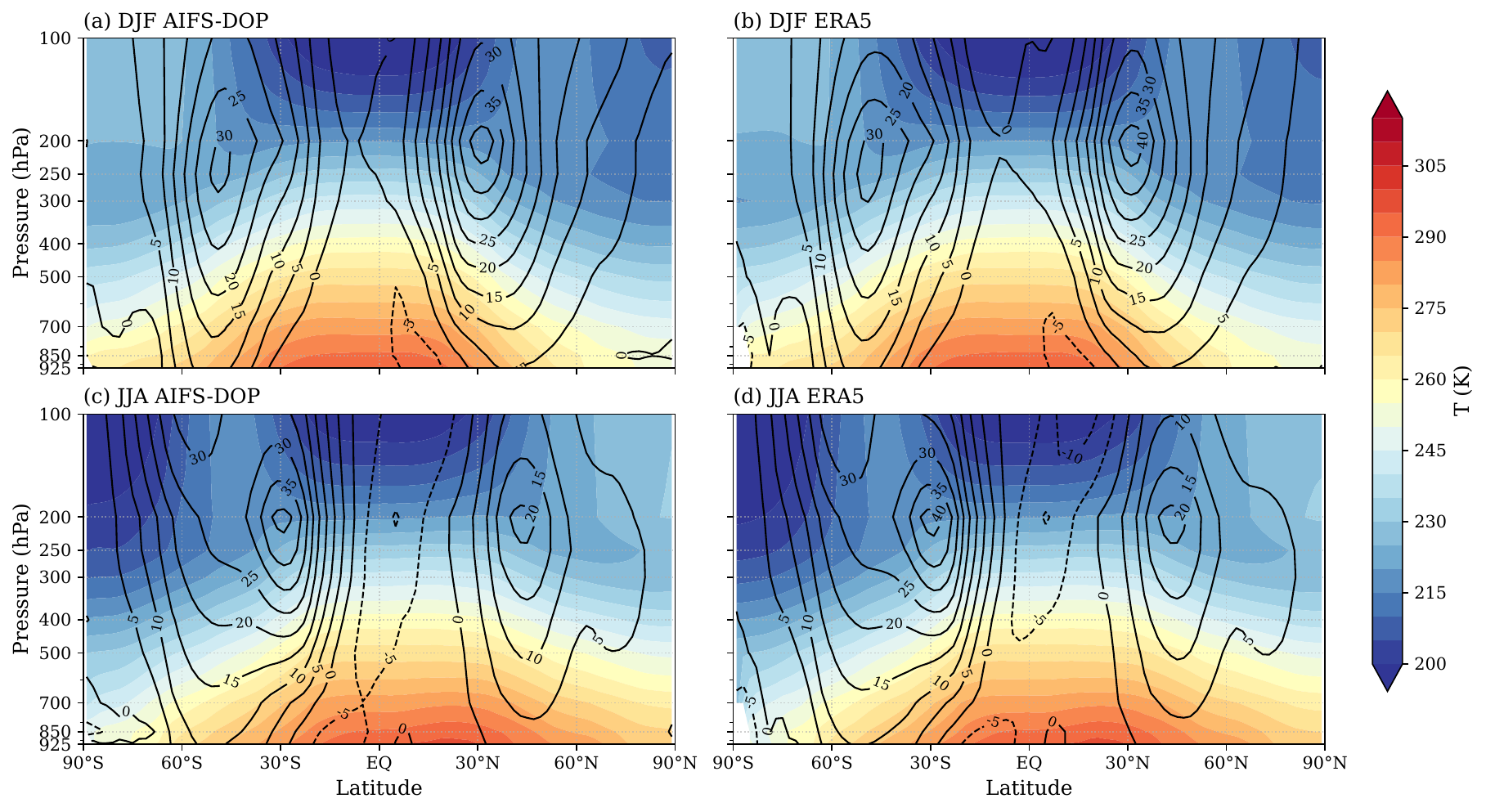}
  \caption{Zonal mean cross-sections of temperature (shaded, K) and zonal wind (black contours, m/s) for the boreal winter (December--February; DJF) and boreal summer (June--August; JJA) seasons. The left column (a, c) shows the AIFS-DOP dataset, while the right column (b, d) shows the ERA5 reanalysis. Data are averaged over a 20-year period from 2000 to 2019. Zonal wind contours are plotted every 5 m/s, with solid lines representing westerly winds and dashed lines representing easterly winds. The vertical axis utilizes a logarithmic pressure scale, and subsurface grid points (where the pressure level exceeds the surface pressure) have been masked.}
  \label{fig:atmospheric-structure-mean-state}
\end{figure}

\begin{figure}[H]
  \centering
  \includegraphics[width=0.75\linewidth]{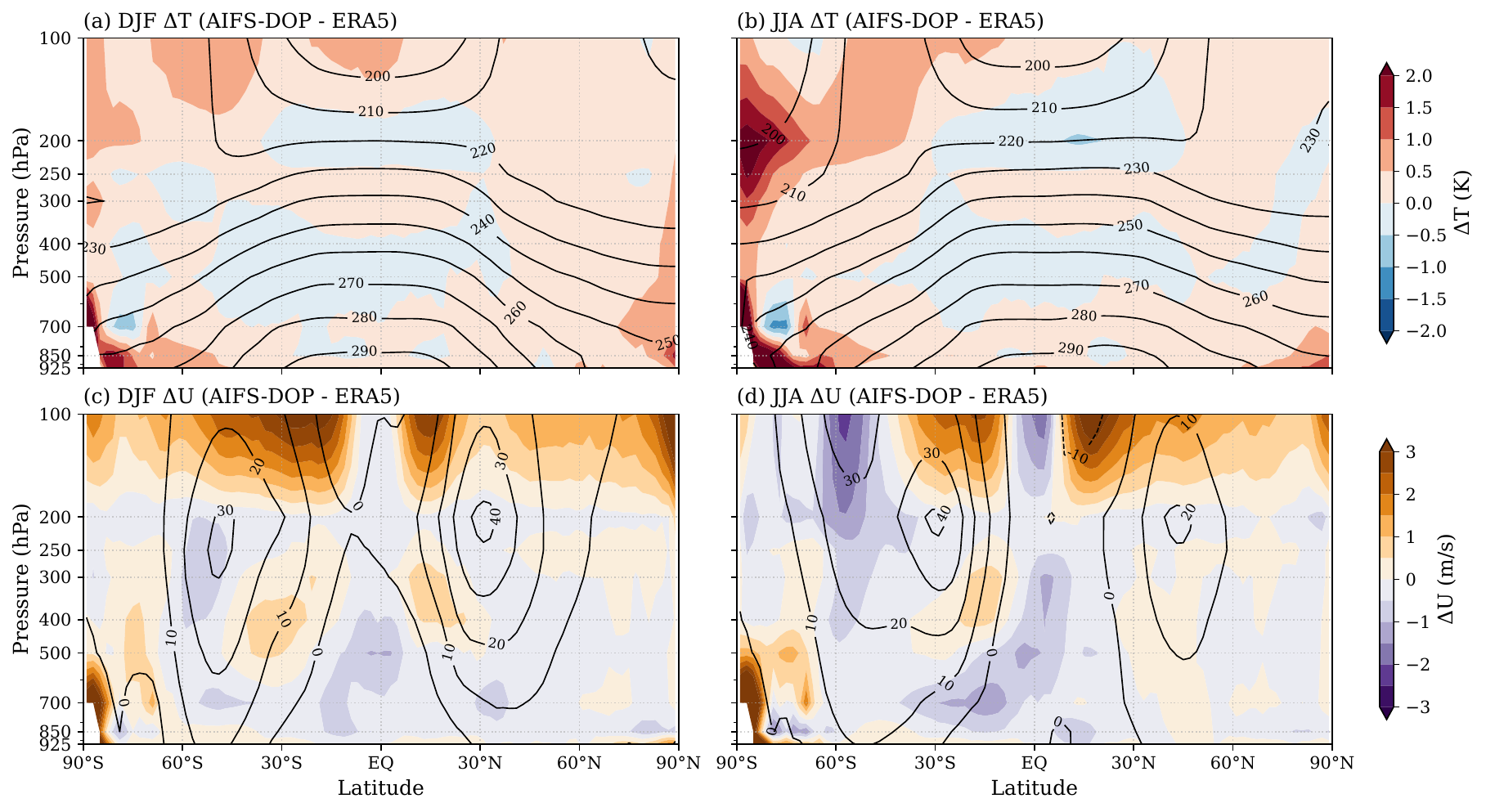}
  \caption{Cross-sections of the zonal mean differences between AIFS-DOP and ERA5 (calculated as AIFS-DOP minus ERA5) averaged over the 2000--2019 period. The top row shows temperature biases ($\Delta T$, shaded, K) for (a) DJF and (b) JJA. The bottom row shows zonal wind biases ($\Delta U$, shaded, m/s) for (c) DJF and (d) JJA. To provide climatological context, the corresponding full ERA5 absolute fields are overlaid as black contours on each panel. Overlay contour intervals are 10 K for temperature and 10 m/s for zonal wind, with negative wind values denoted by dashed lines. Subsurface regions are masked.}
  \label{fig:extended-data-zonal-structure-differences}
\end{figure}

A basic requirement for any reanalysis is that it represents the large-scale mean structure of the atmosphere, including the zonal jets, thermal stratification, and meridional overturning circulation. Zonal-mean cross-sections of zonal wind and temperature for December to February (DJF) and June to August (JJA) are shown in Figure~\ref{fig:atmospheric-structure-mean-state} for both AIFS-DOP and ERA5, while differences between the two datasets are shown in Figure~\ref{fig:extended-data-zonal-structure-differences}. 

Despite the sparsity of the in-situ upper-air observing network, AIFS-DOP aligns closely with ERA5. Absolute differences in the zonal mean temperature field are generally less than 0.5 K. The main exception to this is over Antarctica where AIFS-DOP is warmer than ERA5, particularly in JJA. The subtropical jets are positioned as would be expected from thermal wind relationships with respect to the horizontal temperature gradients, and are well captured in terms of both intensity and location. Similarly, the polar night jet can be seen in JJA at 100~hPa associated with the temperature gradient caused by strong radiative cooling in that region, although its intensity is slightly less than ERA5. Meanwhile, the reversal of the near-surface trade wind easterlies is present with the latitude varying as expected with season. The tropical easterlies at upper levels in JJA are present in AIFS-DOP but extend further towards the surface than in ERA5.

\begin{figure}[H]
  \centering
  \includegraphics[width=0.98\linewidth]{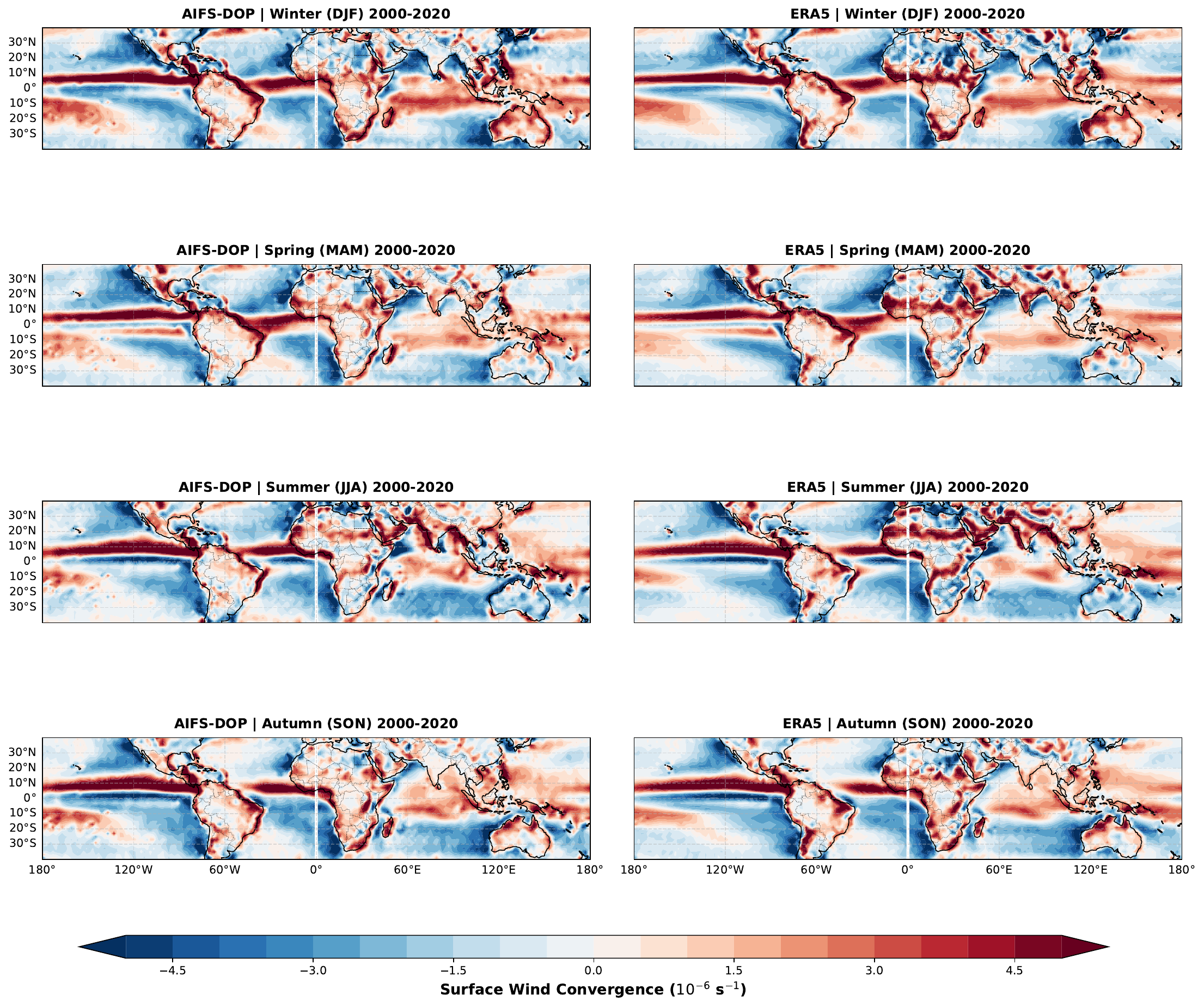}
  \caption{Seasonal climatology of tropical surface wind convergence (2000--2019). 20-year seasonal averages of 10-metre wind convergence for AIFS-DOP (left) and ERA5 (right) restricted to $40^\circ\text{S}$ to $40^\circ\text{N}$. Rows display boreal winter (DJF), spring (MAM), summer (JJA), and autumn (SON). Units are $10^{-6} \text{ s}^{-1}$.}
  \label{fig:extended-data-itcz}
\end{figure}

Figure~\ref{fig:extended-data-itcz} shows the mean convergence of the 10-metre winds for each season. The Inter-Tropical Convergence Zone (ITCZ) is clearly visible and its migration north and south with the changing seasons is well captured in AIFS-DOP. In addition, some finer scale details such as the narrow band of divergence along the equator in the East Pacific during March to May (MAM) associated with the tongue of cold SSTs in this region are visible in both datasets.

Over Africa, the inter-tropical discontinuity convergence line is present at the correct latitude in AIFS-DOP but its intensity is weaker than in ERA5. While the convergence along the equator marks ascent associated with convection, the broad areas of divergence in the sub-tropics indicate the subsidence associated with the descending arms of the Hadley cells. Over the west Pacific, the diagonal line of the South Pacific Convergence Zone shows good agreement with ERA5. Some smaller spots of increased convergence can be seen near islands which we speculate could be artifacts related to the locations of surface stations.

These results are particularly encouraging given that this model was not trained or initialised with scatterometer data; the surface wind predictions were informed only by sparse in-situ wind and pressure observations alongside indirect satellite radiance data.

\begin{figure}[H]
  \centering
  \captionsetup[subfigure]{font=small,skip=2pt}
  \begin{subfigure}{0.9\linewidth}
    \centering
    \includegraphics[width=\linewidth]{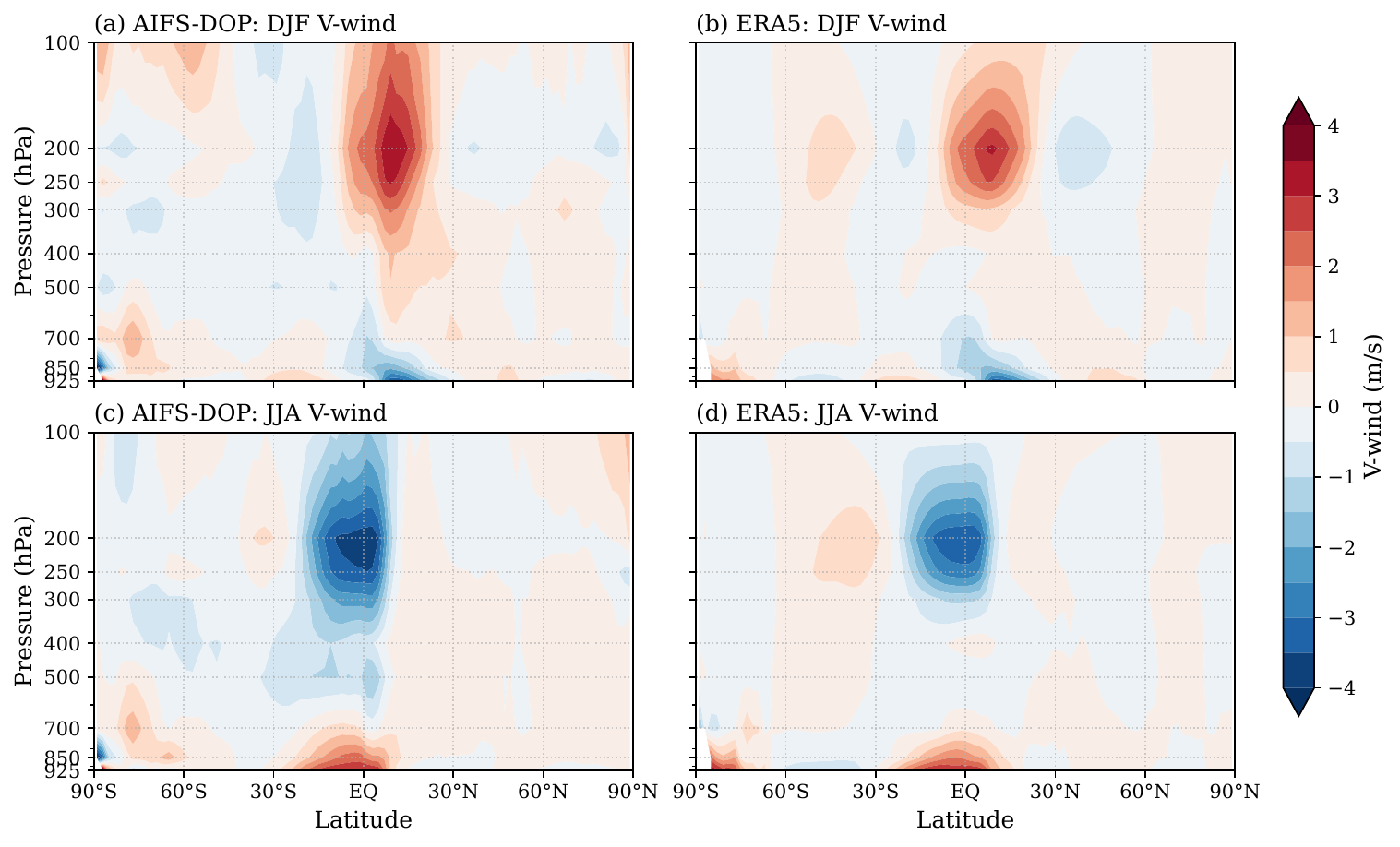}
  \end{subfigure}
  \caption{Zonal mean meridional wind climatology (2000--2019). Cross-sections of the meridional circulation (m/s) for (a, c) AIFS-DOP and (b, d) ERA5 during the (a, b) December--February (DJF) and (c, d) June--August (JJA) seasons. Grid points where the target pressure level exceeds the local surface pressure are excluded.}
  \label{fig:extended-data-jja-atmospheric-structure}
\end{figure}

Moving away from the surface, the zonal mean meridional winds are shown in Figure~\ref{fig:extended-data-jja-atmospheric-structure}. While AIFS-DOP reproduces the low-level convergence and upper-level divergence associated with the Hadley cells there are more marked differences at mid-levels with AIFS-DOP exhibiting deeper areas of convergence and divergence relative to ERA5. We consider this likely to be an area that needs further improvement in AIFS-DOP.

\begin{figure}[H]
  \centering
  \captionsetup[subfigure]{font=small,skip=2pt}
  \begin{subfigure}{0.9\linewidth}
    \centering
    \includegraphics[width=\linewidth]{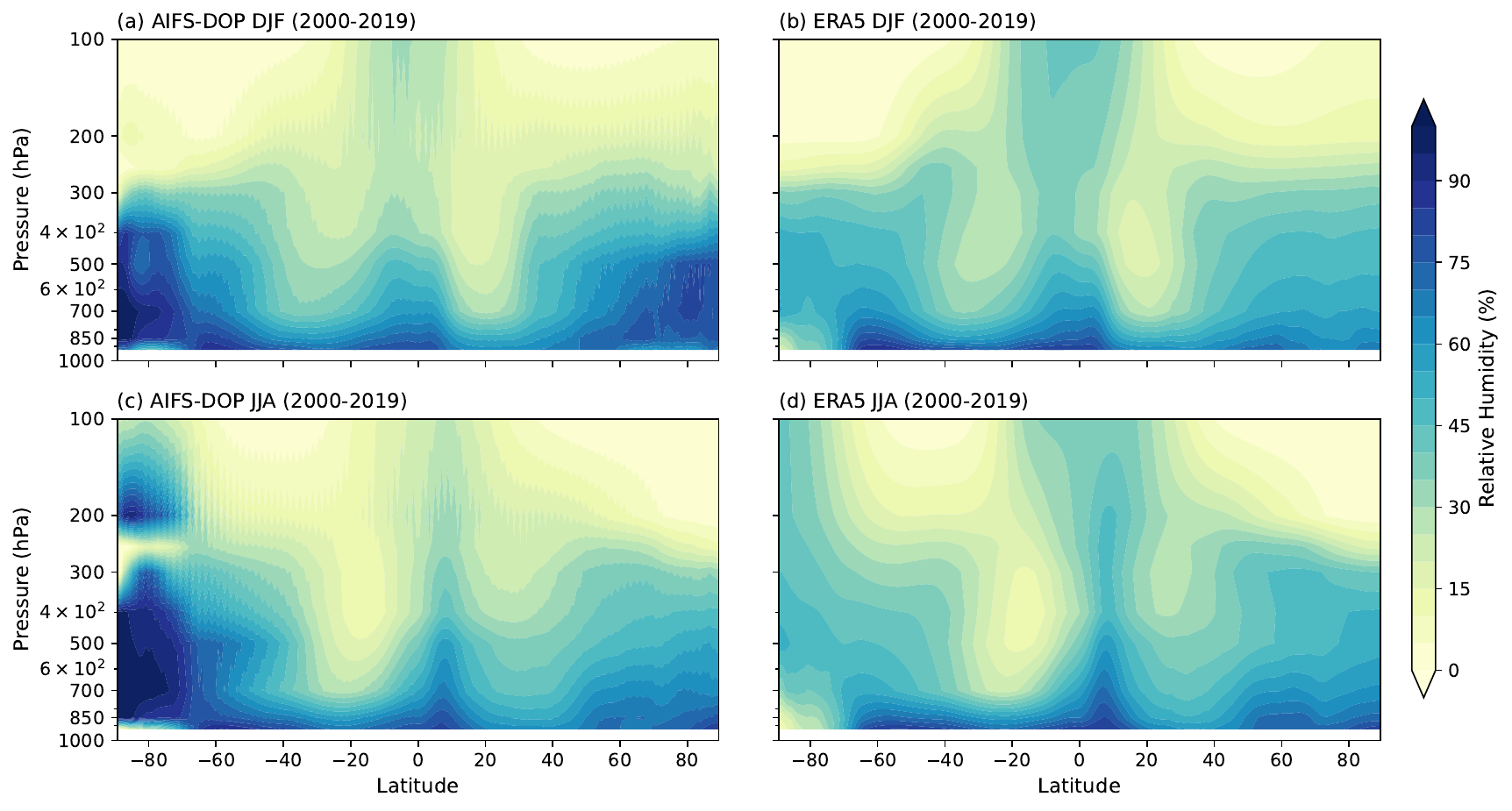}
  \end{subfigure}
  \caption{Zonal mean relative humidity climatology (2000--2019). Cross-sections of the relative humidity (percent) for (a, c) AIFS-DOP and (b, d) ERA5 during the (a, b) December--February (DJF) and (c, d) June--August (JJA) seasons. Grid points where the target pressure level exceeds the local surface pressure are excluded.}
  \label{fig:extended-data-rh-climatology}
\end{figure}

Predictions for zonal mean relative humidity are shown in Figure~\ref{fig:extended-data-rh-climatology}. AIFS-DOP captures several broadscale features found in ERA5. For example, the high tropospheric humidity associated with convection along the ITCZ and the dry air in the descending arms of the Hadley cells are both present. However, this figure also highlights implausibly high humidity values in AIFS-DOP in regions with very dry air such as the poles and the stratosphere. We hypothesise this is related to the fact that the model is trained on specific humidity measurements. Since typical specific humidity values are several orders of magnitude larger at the equator than at the pole, the learning signal from the MSE loss will be dominated by predictions in areas of high humidity. Training against dew point or relative humidity targets instead of specific humidity may improve this as it would tend to increase the relative contribution of humidity mismatches at the poles to the overall loss.

\section*{Multi-scale variability}

The Earth system exhibits variability across a wide range of spatial and temporal scales, and reanalysis datasets have been central to studying and improving our understanding of this variability. An observation-driven reanalysis therefore needs to recover not only the mean atmospheric state, but also representative modes of variability across synoptic, seasonal, and longer timescales. We focus here on a set of illustrative diagnostics: extra-tropical storm tracks, tropical variability, and decadal temperature trends.

\begin{figure}[H]
  \centering
  \captionsetup[subfigure]{font=small,skip=2pt}
  \begin{subfigure}{0.94\linewidth}
    \centering
    \includegraphics[width=\linewidth]{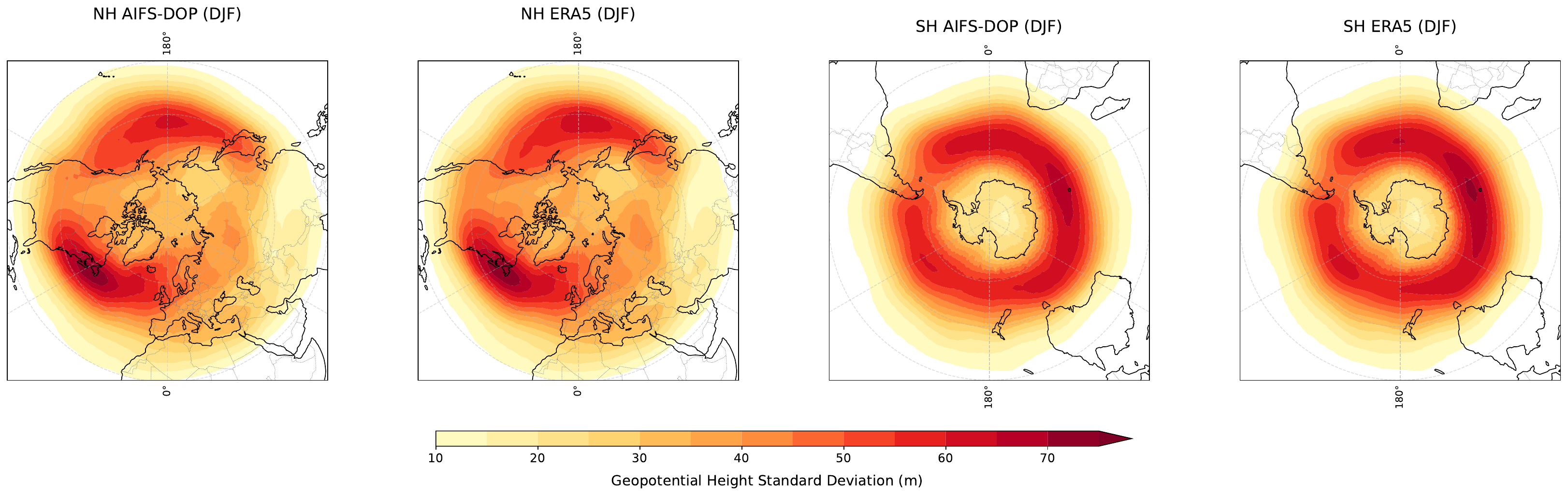}
    \caption{DJF storm tracks in both hemispheres.}
    \label{fig:storm-track-djf}
  \end{subfigure}

  \vspace{0.15em}

  \begin{subfigure}{0.94\linewidth}
    \centering
    \includegraphics[width=\linewidth]{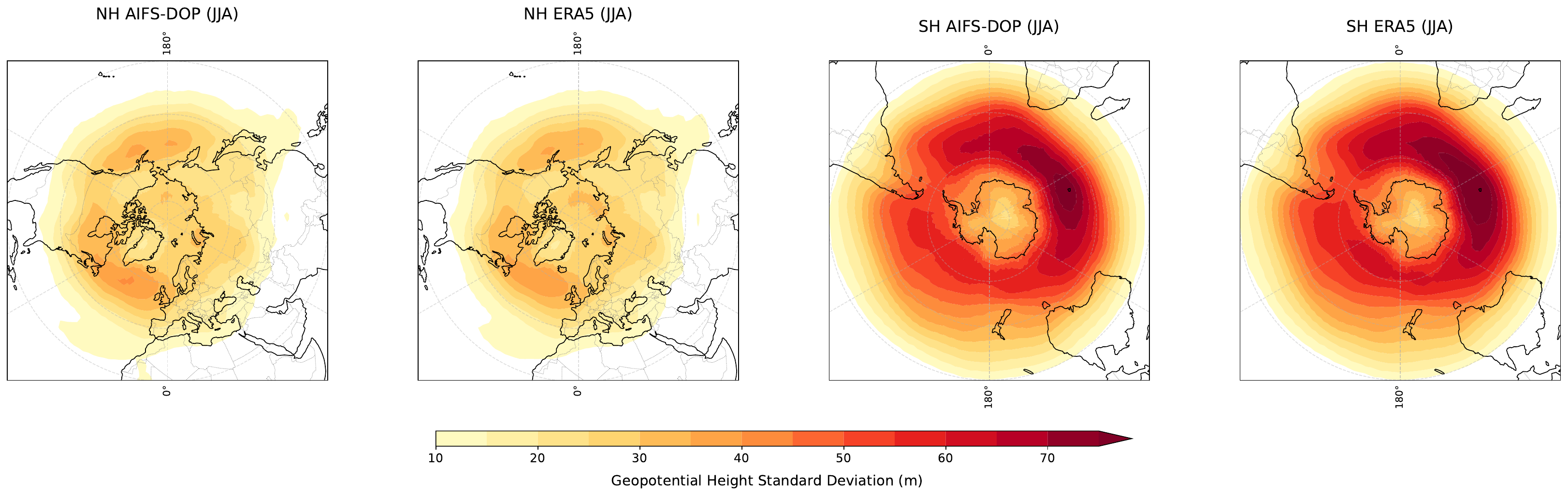}
    \caption{JJA storm tracks in both hemispheres.}
    \label{fig:storm-track-jja}
  \end{subfigure}

  \vspace{0.15em}

  \begin{subfigure}{0.76\linewidth}
    \centering
    \includegraphics[width=\linewidth]{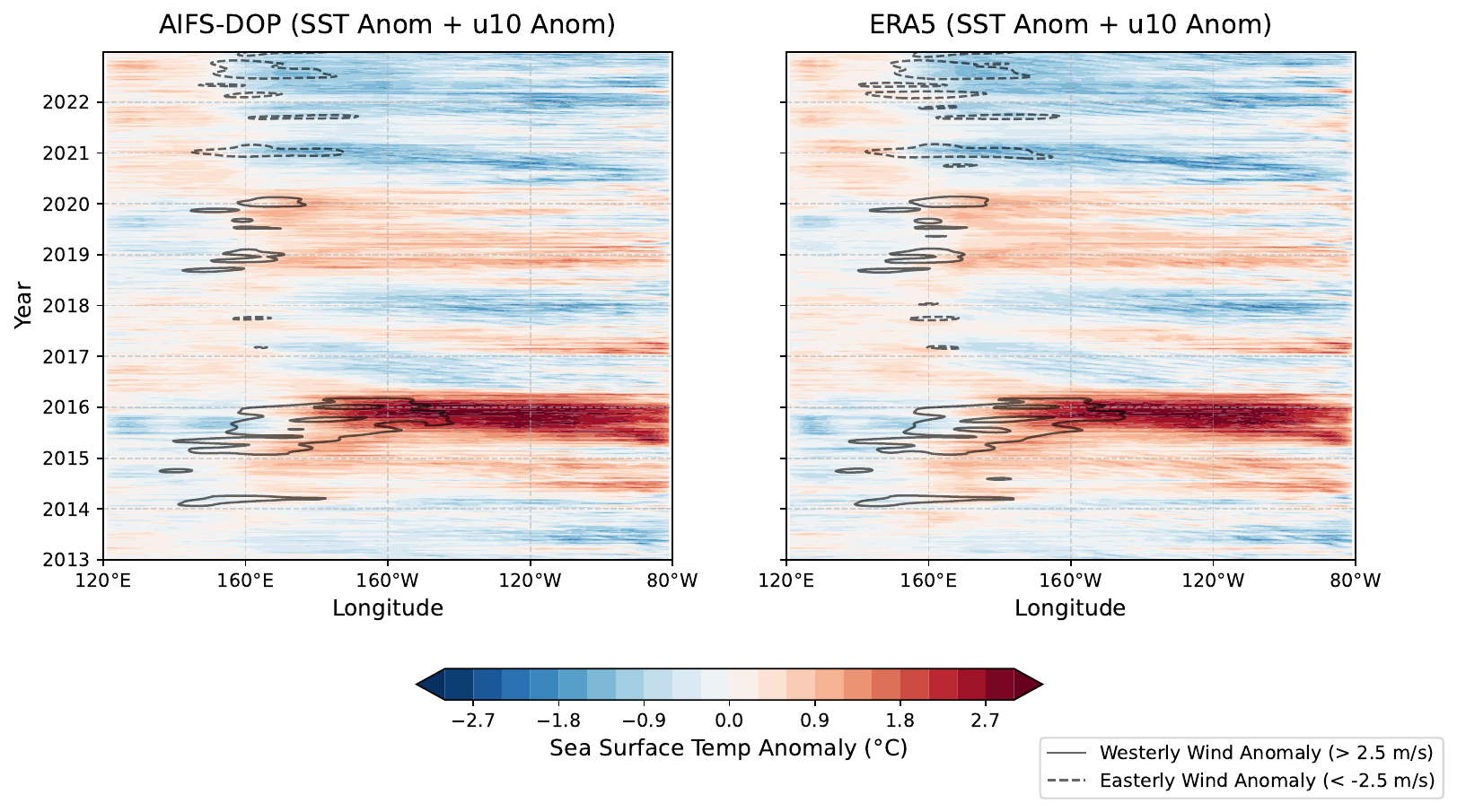}
    \caption{Hovm{\"o}ller diagnostic of ENSO variability.}
    \label{fig:hovmoller-enso}
  \end{subfigure}

  \caption{(a, b) Extratropical storm-track structure for the Northern and Southern Hemispheres during (a) DJF and (b) JJA. Storm tracks are quantified as the standard deviation of the 2--6 day bandpass-filtered 500~hPa geopotential height in metres with statistics computed from 2002 to 2021.
(c) Hovmöller diagram of tropical ENSO variability over the equatorial Pacific (between $5^\circ$S and $5^\circ$N) for the period 2013--2022. Coloured shading denotes weekly SST anomalies ($^\circ$C) with overlaid contours for 10-metre zonal wind anomalies. All anomalies are computed relative to the monthly mean of the plotted period.}
  \label{fig:multi-scale-variability}
\end{figure}

Figures~\ref{fig:storm-track-djf} and \ref{fig:storm-track-jja} show the standard deviation of the 2-6 day band-pass filtered geopotential height at 500~hPa as an indicator of extra-tropical storm track intensity. The maps show that AIFS-DOP captures the spatial distribution of the transient eddy variance well, as well as the change in intensity with season. This indicates that the model is producing baroclinic eddies in the right locations with approximately the correct intensity. In the southern hemisphere the storm tracks are slightly weaker than in ERA5. 

The tropical variability of AIFS-DOP is explored in Figure~\ref{fig:hovmoller-enso}. The observations used to train and initialise AIFS-DOP included sea surface temperature (SST) observations from buoys as well as indirect satellite radiances with surface sensitivities. The inter-annual tropical SST variability associated with El Ni\~no/Southern Oscillation (ENSO) is well captured with the strong El Ni\~no of 2015-2016 dominating the plot. In addition, the zonal wind anomalies associated with the shifted Walker circulation in response to ENSO can be clearly seen.

\begin{figure}[H]
  \centering
  \includegraphics[width=0.98\linewidth]{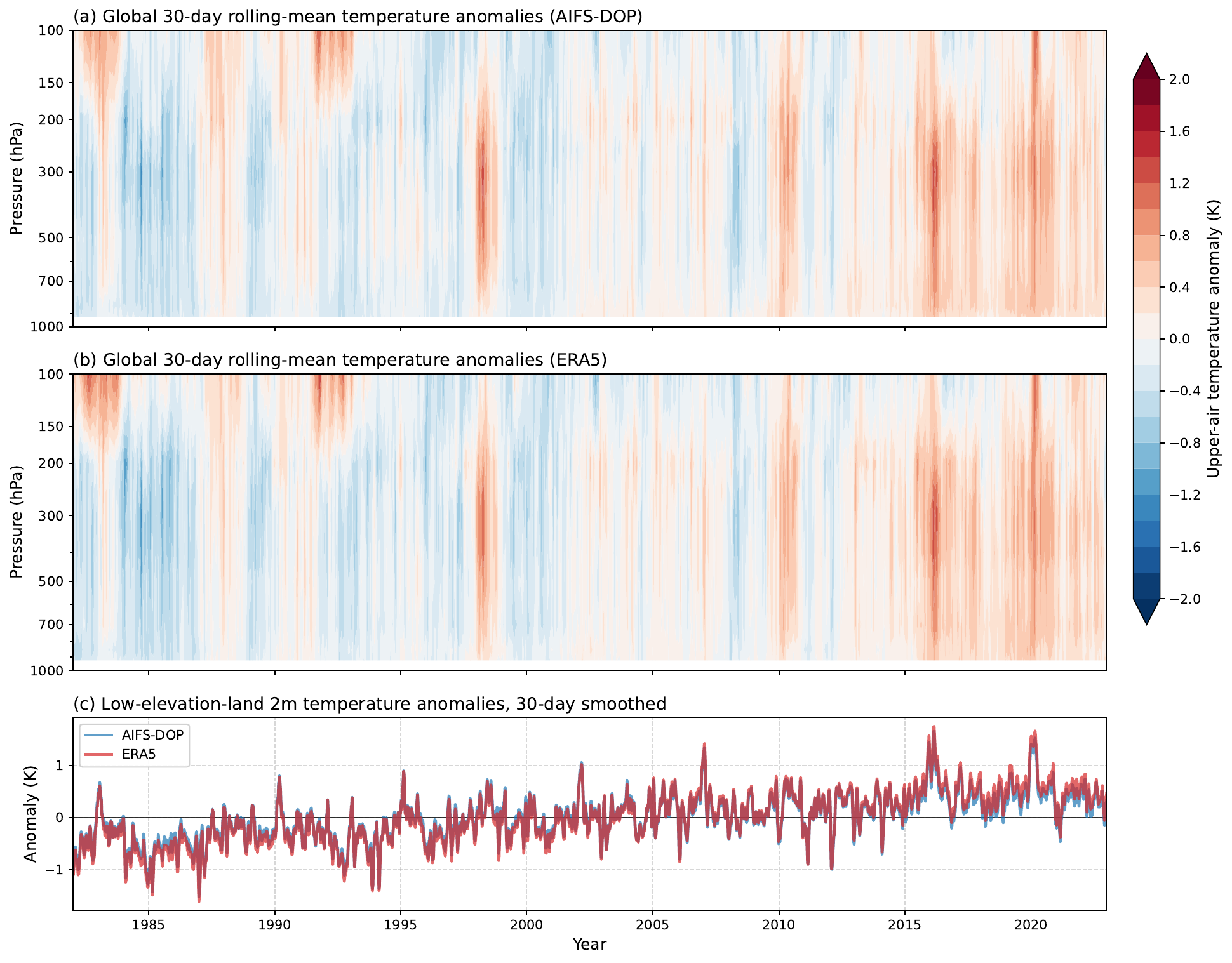}
  \caption{Global temperature anomalies comparing the AIFS-DOP reanalysis with ERA5.1 (1982--2022). (a, b) Time-pressure cross-sections of global upper-air temperature anomalies for (a) AIFS-DOP and (b) ERA5.1. Anomalies are calculated by subtracting a daily climatology from the daily means, followed by a centered 30-day rolling mean. Coloured shading representing the temperature anomaly in Kelvin (K). (c) Time series of 2-metre temperature anomalies over low-elevation (<200 metre) land areas for AIFS-DOP (blue line) and ERA5 (red line). These anomalies are computed relative to a daily climatology and smoothed using a centered 30-day rolling mean.}
  \label{fig:temperature-anomaly-timeseries}
\end{figure}

As an example of variability on decadal timescales, Figure~\ref{fig:temperature-anomaly-timeseries} shows the upper-air (a, b) and near-surface (c) temperature anomalies. Several features of interest are worth highlighting; stratospheric warmings associated with the volcanic eruptions of El Chichon in 1982 and Pinatubo in 1991 are clearly visible in both datasets. In the troposphere, prominent warming associated with El Ni\~no events (in particular the 1997 event) and cooling associated with La Ni\~na can be seen. On longer timescales, the gradual global tropospheric warming signal is evident. Since this model was not trained with radiosonde data above 100~hPa it is not possible to see the cooling associated with increased CO2 in the stratosphere. At the surface, there is a high degree of correlation between the monthly two-metre temperature anomalies over low elevation land in the two datasets. In addition, a clear signal of global warming is present. High elevation land points were excluded here to avoid complexities when comparing two models with different native resolutions in areas of complex topography since 2m temperature has a strong dependence on height. 

\begin{figure}[H]
  \centering
  \captionsetup[subfigure]{font=small,skip=2pt}
  \begin{subfigure}{0.76\linewidth}
    \centering
    \includegraphics[width=\linewidth]{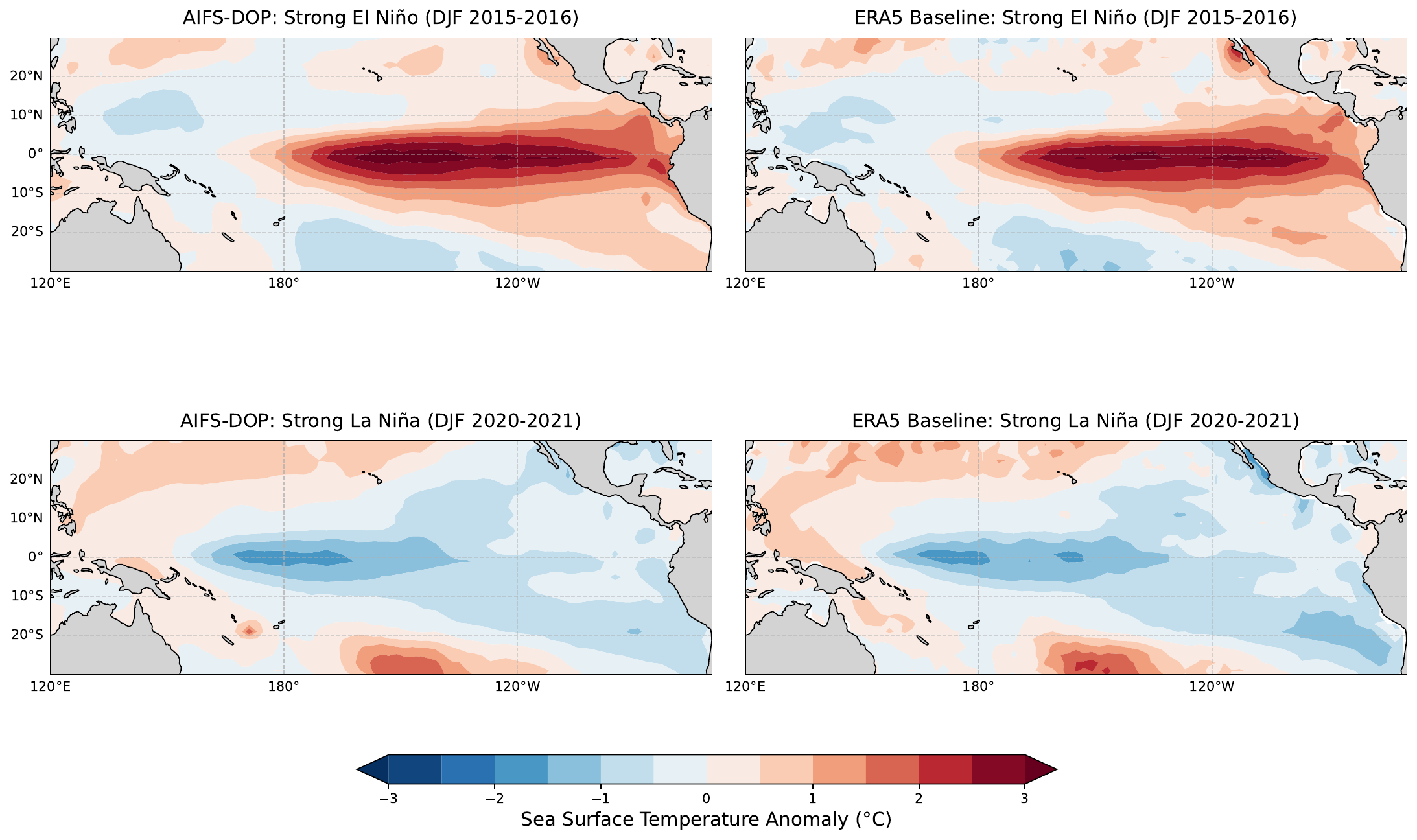}
    \caption{SST anomaly composite.}
    \label{fig:enso-sst-anomaly}
  \end{subfigure}

  \vspace{0.4em}

  \begin{subfigure}{0.72\linewidth}
    \centering
    \includegraphics[width=\linewidth]{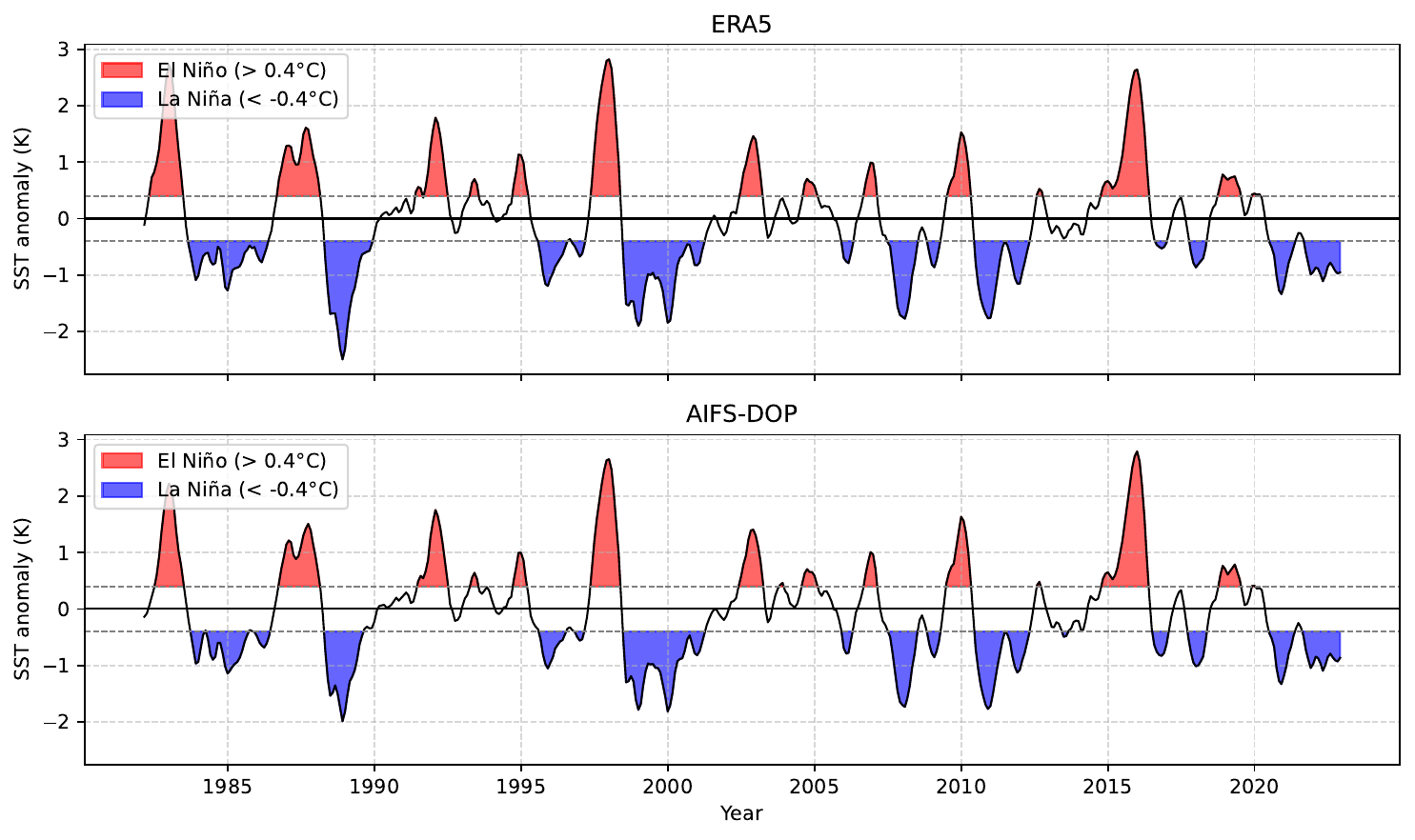}
    \caption{Ni\~{n}o3.4 time series.}
    \label{fig:enso-nino34-timeseries}
  \end{subfigure}

  \caption{El Niño--Southern Oscillation (ENSO) sea surface temperature (SST) anomalies. (a) Boreal winter (DJF) SST anomalies ($^\circ$C) during the strong 2015--2016 El Niño (top) and 2020--2021 La Niña (bottom) events for AIFS-DOP (left) and ERA5 (right), relative to the 2000--2020 DJF baseline. (b) Oceanic Niño Index (ONI) time series (1982--2022) for the Niño 3.4 region ($5^\circ$N--$5^\circ$S, $120^\circ$W--$170^\circ$W). Time series display the 3-month centered rolling mean of SST anomalies (K) relative to a 1991--2020 monthly climatology. Red and blue shading highlight El Niño (> 0.4 K) and La Niña (< -0.4 K) periods, respectively.}
  \label{fig:extended-data-enso-sst-nino34}
\end{figure}

The spatial (Figure~\ref{fig:enso-sst-anomaly}) and temporal (Figure~\ref{fig:enso-nino34-timeseries}) distribution of the tropical SST anomalies shows close alignment with ERA5. Some of the anomalies appear slightly smoothed compared to ERA5. In particular towards the beginning of the period the 1988/9 La Ni\~na appears weaker in AIFS-DOP than ERA5.

\begin{figure}[H]
  \centering
  \includegraphics[width=0.92\linewidth]{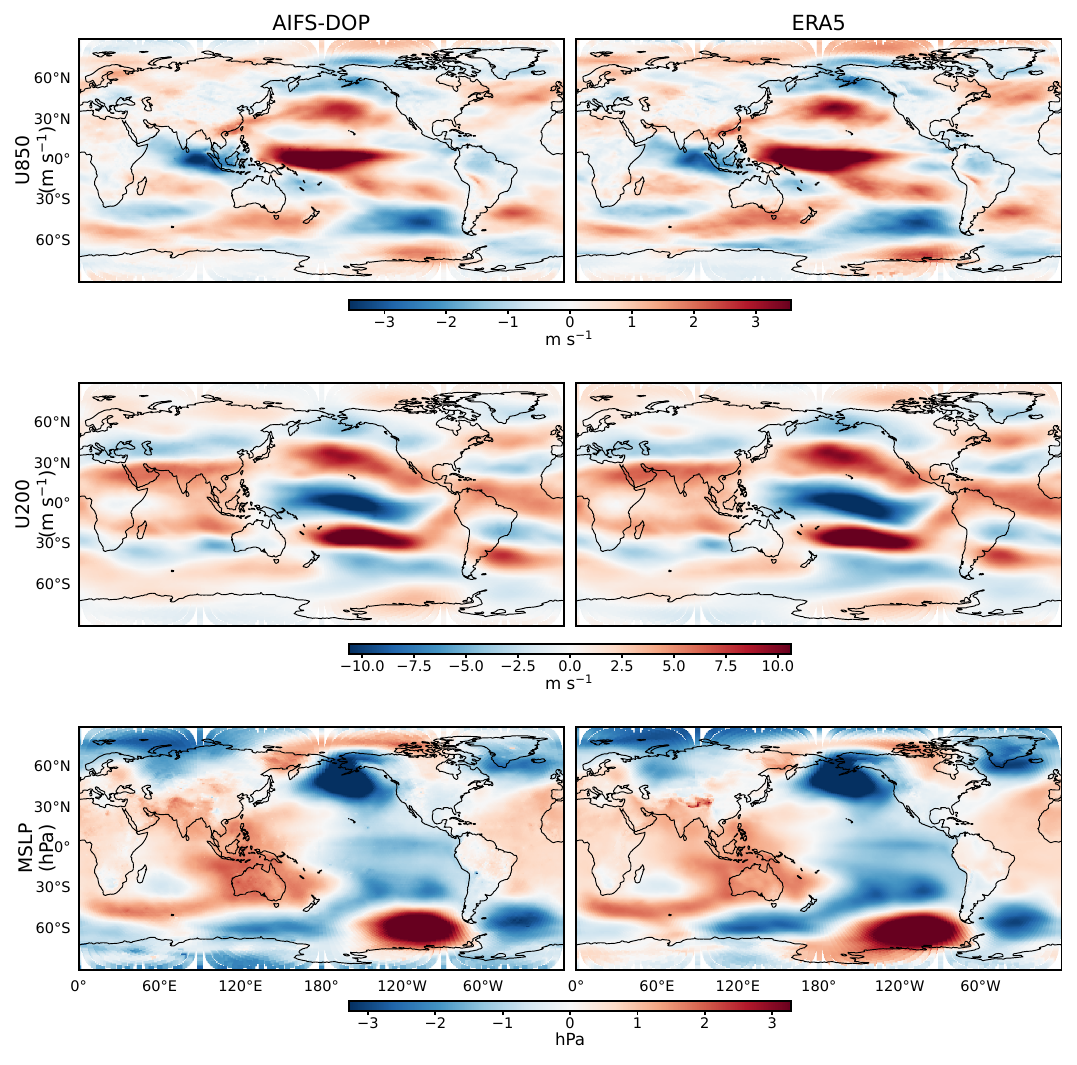}
  \caption{El Niño minus La Niña atmospheric composites (2013--2022). Global composite differences between El Niño and La Niña conditions for AIFS-DOP (left column) and ERA5 (right column). Rows display the difference in anomalies for 850~hPa zonal wind (top, m s$^{-1}$), 200~hPa zonal wind (middle, m s$^{-1}$), and mean sea level pressure (bottom, hPa). ENSO phases are defined using a $\pm$0.5 K threshold applied to the 3-month centered rolling mean of the Niño 3.4 sea surface temperature anomaly index.}
  \label{fig:extended-data-enso-teleconnections}
\end{figure}

AIFS-DOP also captures the teleconnection patterns associated with ENSO. Figure \ref{fig:extended-data-enso-teleconnections} shows composite difference plots of the anomalies between El Ni\~no and La Ni\~na events. The weakening of the Walker circulation is highlighted by the reduced zonal winds along the equator at 850~hPa and the corresponding slowdown at 200~hPa. Furthermore, the teleconnection patterns away from the tropics at upper levels with the strengthened sub-tropical jets and in the mean sea level pressure in the extra-tropics show remarkable similarity to the corresponding patterns in ERA5.

\begin{figure}[H]
  \centering
  \captionsetup[subfigure]{font=scriptsize,skip=1pt}
  \begin{subfigure}{0.58\linewidth}
    \centering
    \includegraphics[width=\linewidth]{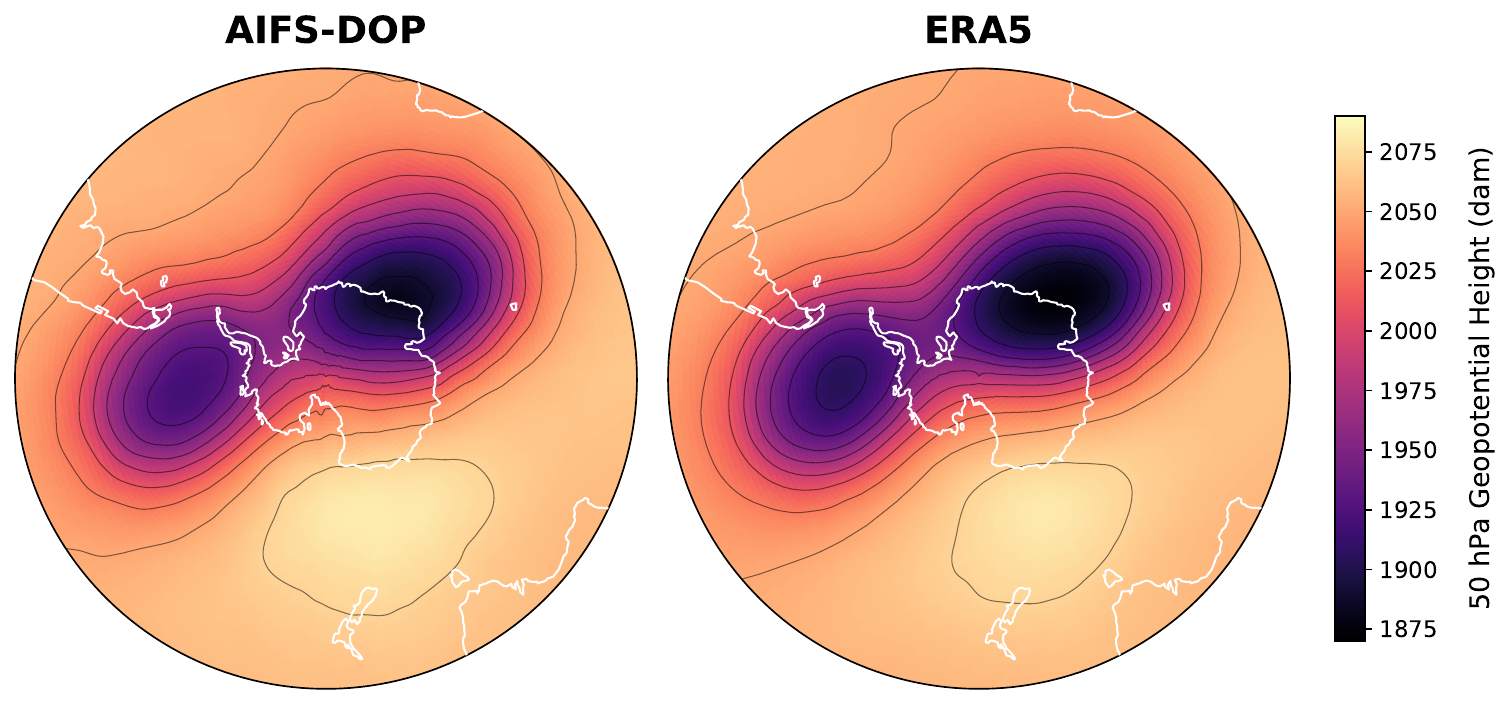}
    \caption{Stratospheric polar-vortex splitting case study.}
    \label{fig:polar-vortex-split-case-study}
  \end{subfigure}

  \vspace{0.1em}

  \begin{subfigure}{0.58\linewidth}
    \centering
    \includegraphics[width=\linewidth]{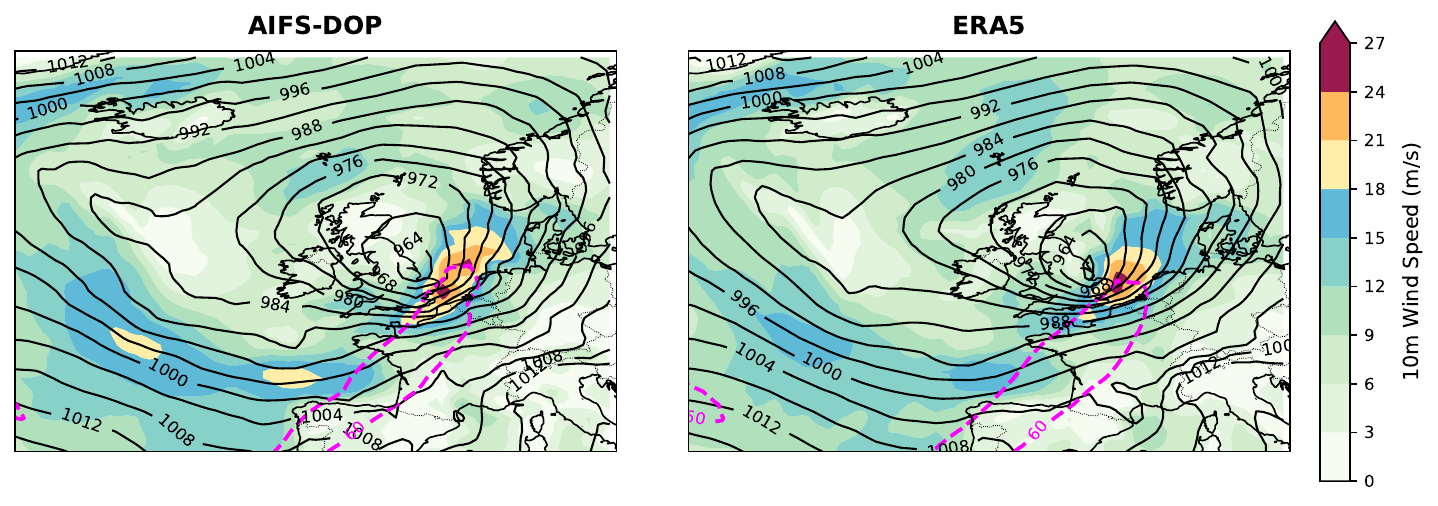}
    \caption{1987 storm case study.}
    \label{fig:storm-1987-case-study}
  \end{subfigure}

  \vspace{0.1em}

  \begin{subfigure}{0.58\linewidth}
    \centering
    \includegraphics[width=\linewidth]{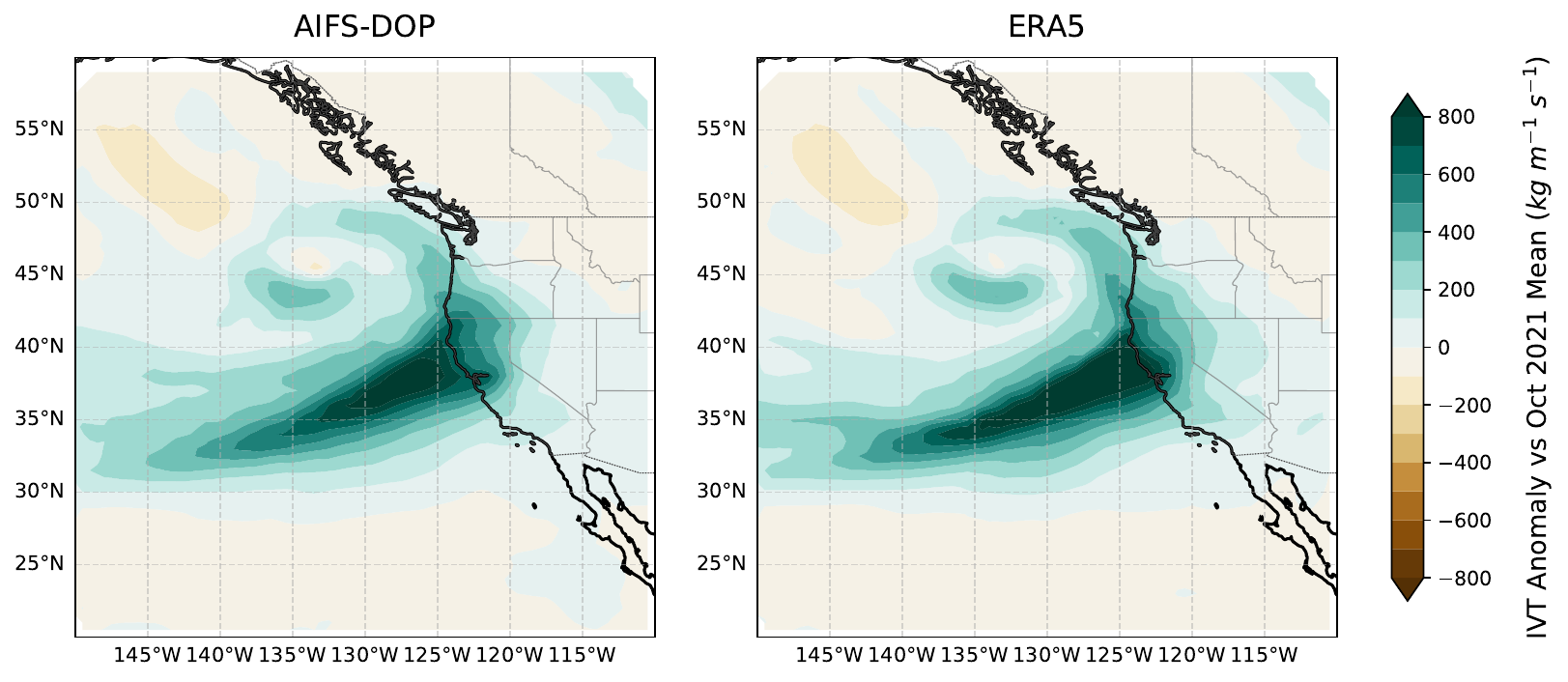}
    \caption{Atmospheric-river case study.}
    \label{fig:atmospheric-river-case-study}
  \end{subfigure}

  \vspace{0.1em}

  \begin{subfigure}{0.58\linewidth}
    \centering
    \includegraphics[width=\linewidth]{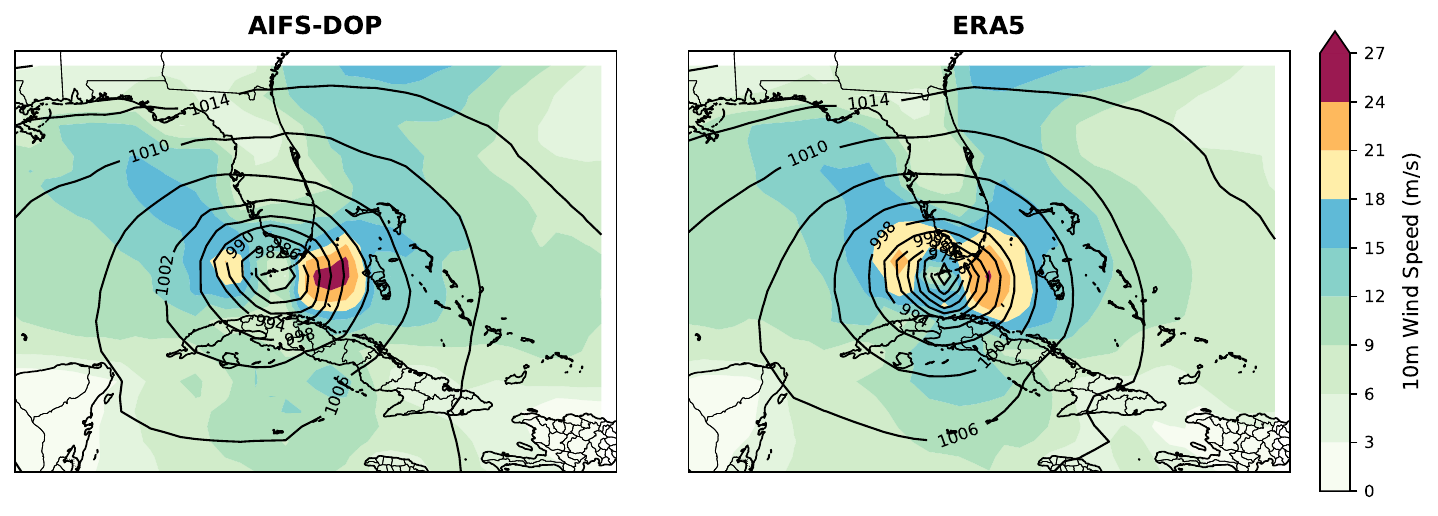}
    \caption{Hurricane Irma case study.}
    \label{fig:hurricane-irma-case-study}
  \end{subfigure}

  \caption{Extreme event case studies comparing AIFS-DOP (left panels) and ERA5 (right panels). (a) Antarctic polar vortex split on 26 September 2002 at 00:00 UTC, displaying 50~hPa geopotential height (dam). (b) The 1987 Great Storm on 16 October 1987 at 06:00 UTC. Shading represents 10m wind speed (m/s), solid black contours denote mean sea level pressure (MSLP, hPa), and dashed magenta contours highlight the 300~hPa jet streak wind speeds (m/s). (c) Atmospheric river on 24 October 2021 at 12:00 UTC. Shading indicates the integrated vapor transport (IVT) anomaly (kg m$^{-1}$ s$^{-1}$) relative to the October 2021 monthly mean, integrated from 925~hPa to 300~hPa. (d) Hurricane Irma at Florida landfall on 10 September 2017 at 12:00 UTC. Fields plotted are identical to those in panel (b).}
  \label{fig:extended-data-case-studies}
\end{figure}

To supplement these large-scale measures of variability, several case studies sampling more extreme events are shown in Figure~\ref{fig:extended-data-case-studies}. For example, the rare Antarctic stratospheric polar vortex split of 2002 was well captured, as was the "Great Storm" of 1987 (an encouraging sign, given that satellite observation coverage in this period was not as complete as in more recent times). AIFS-DOP shows the strong integrated moisture flux marking an atmospheric river event in 2021 (Fig.~\ref{fig:atmospheric-river-case-study}); however, the strength of the moisture anomaly is slightly weaker in the machine learning model. Tropical cyclones lie at the edge of what can be resolved by coarse grid models such as the one presented in this paper (which uses an O96 input and processor mesh - approximately 112~km resolution). Nevertheless, Hurricane Irma (Fig.~\ref{fig:hurricane-irma-case-study}) is represented albeit with the intensity significantly underestimated.

\section*{Physical consistency}
\label{sec:physical_consistency}

A central question of this paper is whether machine learning models trained directly from Earth-system observations, without the explicit physical constraints provided by a numerical forecast model, can generate physically coherent dense gridded fields.

\begin{figure}[H]
  \centering
  \captionsetup[subfigure]{font=small,skip=2pt}
  \begin{subfigure}{0.92\linewidth}
    \centering
    \includegraphics[width=\linewidth]{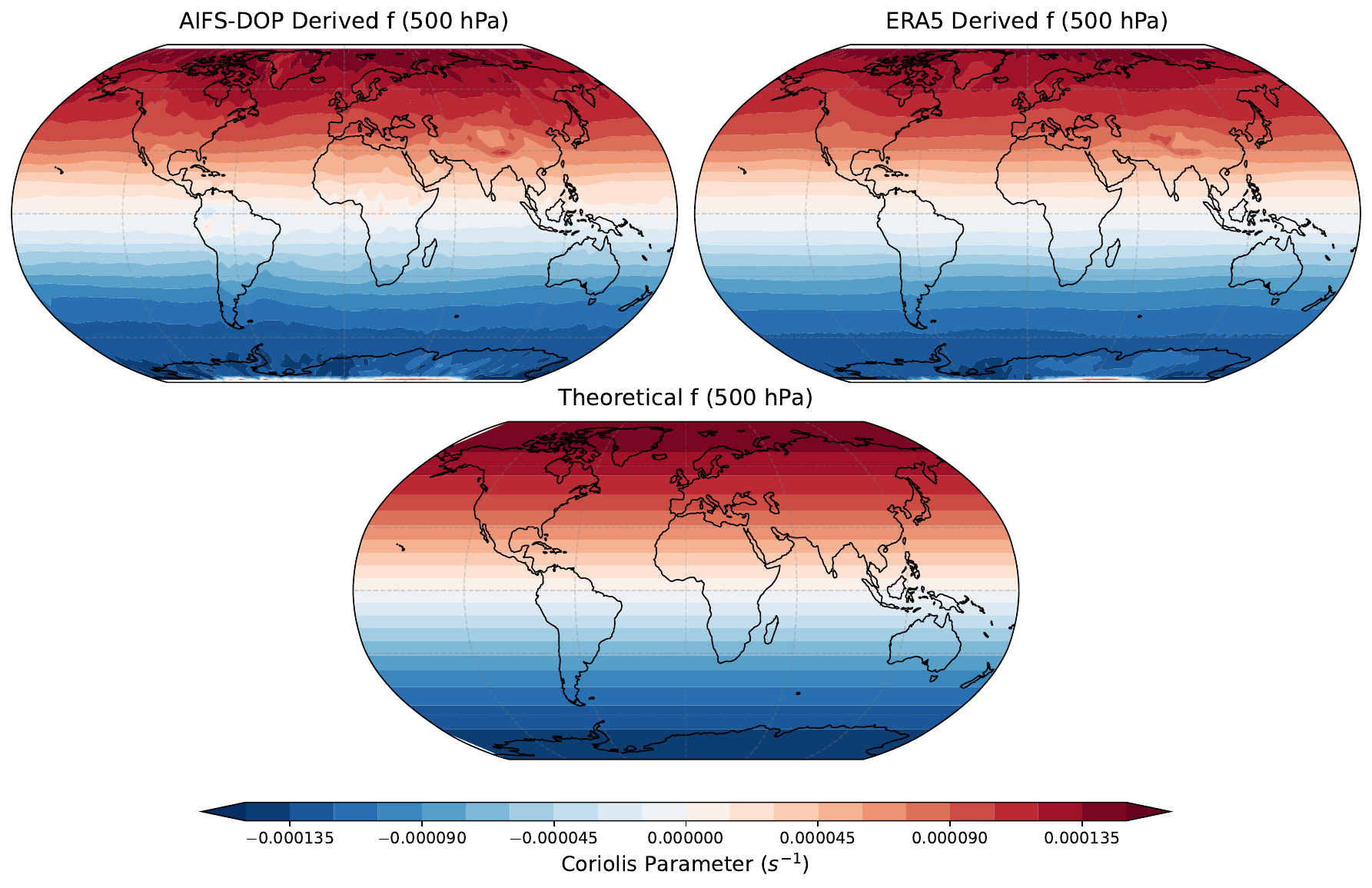}
    \caption{Effective coriolis parameter diagnostic.}
    \label{fig:coriolis-parameter}
  \end{subfigure}

  \vspace{0.35em}

  \begin{subfigure}{0.98\linewidth}
    \centering
    \includegraphics[width=\linewidth]{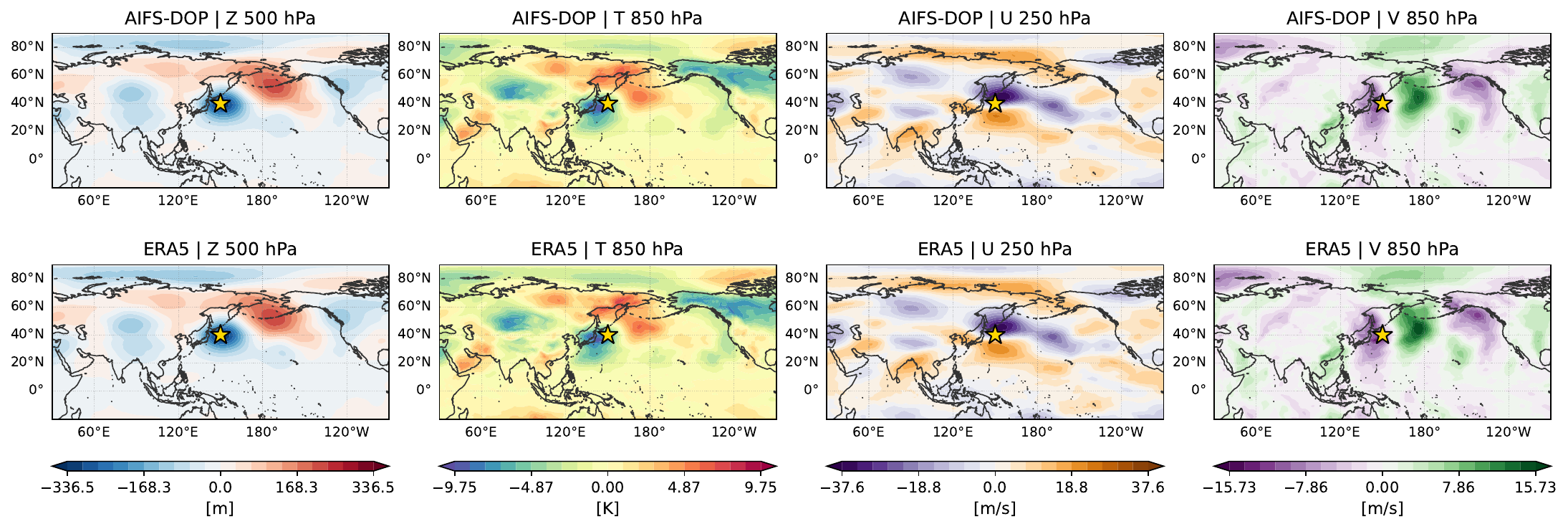}
    \caption{Cross-variable regression patterns associated with a Z500 anomaly.}
    \label{fig:regression-maps}
  \end{subfigure}

  \caption{Physical-consistency diagnostics; (a) time-mean (2000--2019) effective Coriolis parameter at 500~hPa derived from geopotential height and wind fields for AIFS-DOP (left) and ERA5 (right) and compared against the theoretical parameter ($2\Omega\sin(\text{lat})$) (lower panel). Units are s$^{-1}$. (b) Cross-variable regression patterns associated with a z500 anomaly for AIFS-DOP (top row) and ERA5 (bottom row) during the winter of 2017--2018. The fields are calculated by regressing each atmospheric variable against the normalized 500~hPa geopotential height at a reference point in the western North Pacific ($40^\circ$N, $150^\circ$E; marked by a yellow star). Columns, from left to right, display the spatial patterns for 500~hPa geopotential height (m), 850~hPa temperature (K), 250~hPa zonal wind (m/s), and 850~hPa meridional wind (m/s).}
  \label{fig:physical-consistency}
\end{figure}

One important aspect is the relationship between the wind and the mass fields. To first order the winds in the atmosphere are in geostrophic balance with the Coriolis force balancing the pressure gradient force in the absence of friction. Figure \ref{fig:coriolis-parameter} shows a new diagnostic - the \textit{effective Coriolis parameter} implied by the wind and geopotential predictions. Details of its derivation can be found in the Methods section. When this diagnostic is calculated on ERA5 fields (upper-right) it can be seen that the broad change in Coriolis parameter with latitude is recovered with minor differences resulting from ageostrophic dynamics in the atmosphere. The results from AIFS-DOP (upper-left) show a similar pattern; this implies that AIFS-DOP has learned about the changing strength of the Coriolis force with latitude and produces wind and geopotential fields consistent with that. In general the derived Coriolis parameter is noisier in AIFS-DOP, perhaps indicating the presence of grid scale artifacts which remain after time aggregation.

With the aim of uncovering spatial and cross-variable relationships in the gridded fields a linear regression was performed against the 500~hPa geopotential height anomaly at a target grid box in the Pacific, for each grid point and variable, following the approach of Hakim and Masanam (2024) \cite{hakim2024}. In other words, we ask, when there is a strong trough at one location, do the surrounding temperature and wind fields look dynamically consistent with that trough? Figure \ref{fig:regression-maps} plots the fields predicted via these linear relationships associated with a trough at the location marked by the star. The Rossby wave pattern of the trough surrounding the target anomaly (marked by the star), followed by a downstream ridge and a further trough over western Canada, is present in both ERA5 and AIFS-DOP. Similarly, the temperature and wind anomalies associated with the troughs and ridges are reproduced well in AIFS-DOP. This implies that AIFS-DOP is producing fields that are coherent both spatially and across variables and that reproduce the same patterns found in the ERA5 dataset, despite only being trained on sparse observations with no physical model prior.

\section*{Spectral characteristics of the reanalysis fields}

\begin{figure}[H]
  \centering
  \captionsetup[subfigure]{font=small,skip=2pt}
  \begin{subfigure}{0.95\linewidth}
    \centering
    \includegraphics[width=\linewidth]{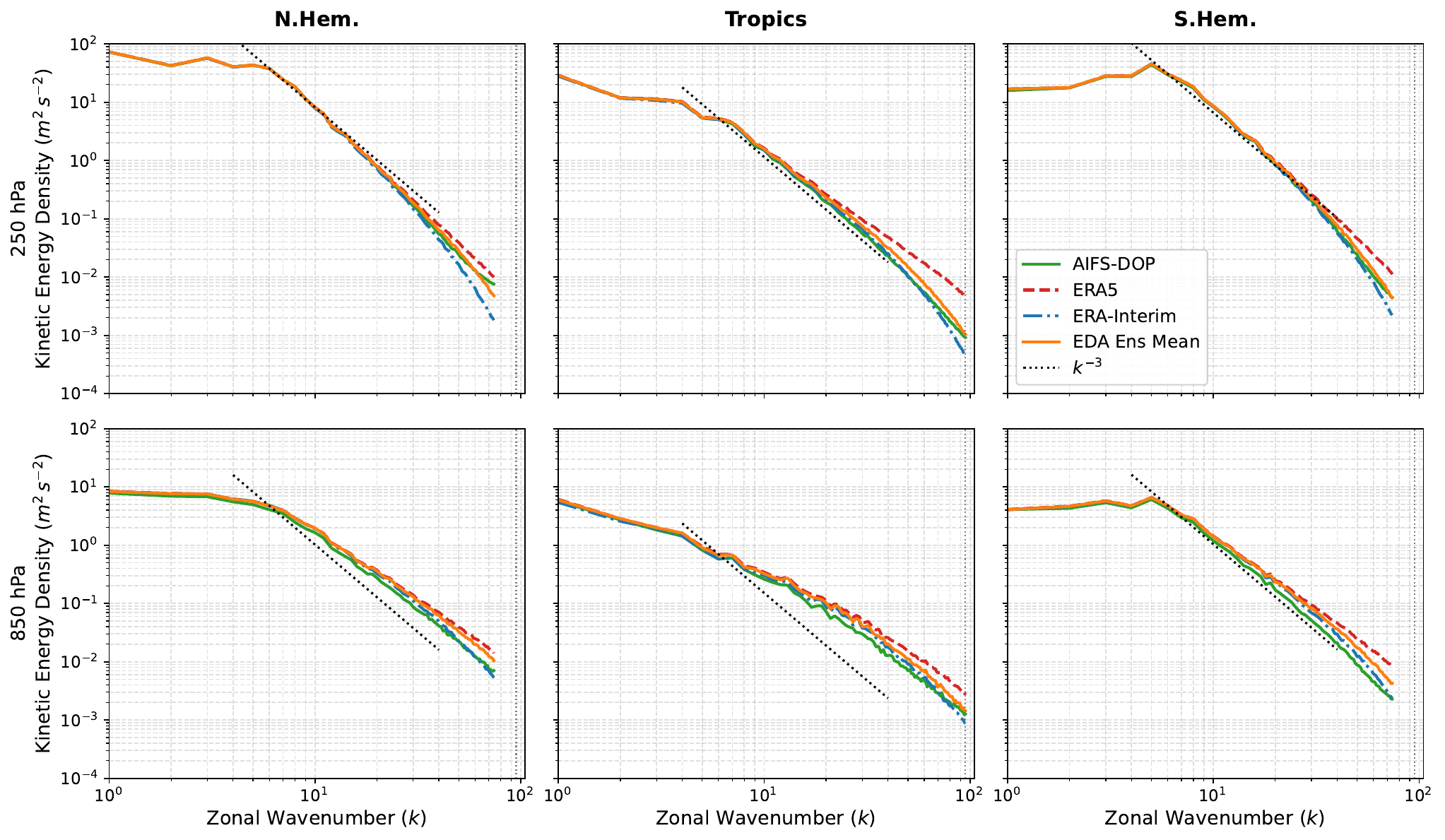}
  \end{subfigure}
  \caption{One-dimensional zonal spectra of horizontal kinetic energy at 250~hPa (top row) and 850~hPa (bottom row) for the northern hemisphere (left), tropics (centre) and southern hemisphere (right). Lines shown for ERA5 at O96 resolution (dashed red), the ERA5 EDA ensemble mean (orange), ERA-Interim at O96 resolution (dashed blue) and AIFS-DOP at O96 (green). Dotted lines show a $k^{-3}$ slope for reference.}
  \label{fig:ke-spectra}
\end{figure}

Figure~\ref{fig:ke-spectra} shows kinetic energy spectra in the east-west direction comparing AIFS-DOP against ERA5, as well as the ERA5 Ensemble of Data Assimilations (EDA) ensemble mean and ERA-Interim. At 250~hPa, it can be seen that there is good agreement in all regions up to around wavenumber 25. At smaller scales AIFS-DOP shows a reduction in energy compared to ERA5; this is true especially in the tropics. We can hypothesise several reasons for this drop-off in energy at small scales.

 Firstly, the low-resolution of the AIFS-DOP prototype (O96, approximately 112~km) is lower than the native resolution of ERA5 (TL639, approximately 31~km). ERA-Interim provides a useful additional reference as its native resolution (TL255, approximately 78~km) is more comparable to AIFS-DOP. Indeed the kinetic energy spectra for ERA-Interim appear to be rather similar to AIFS-DOP.

Secondly, since AIFS-DOP was trained with a mean squared error (MSE) loss, it will tend to express uncertainty by blurring out unpredictable or unobserved scales. Although physical models spin up realistic scales of motion through the turbulent dynamics emerging from Navier-Stokes, the presence of these scales in an analysis does not mean that they are necessarily constrained by observations. Comparison against the EDA ensemble mean provides an indication of the extent to which analysis uncertainty could be contributing to the smoothing seen in AIFS-DOP. Interestingly, at 250~hPa the AIFS-DOP spectra align rather closely with the EDA mean. However at 850~hPa AIFS-DOP shows a closer fit to the ERA-Interim spectra.

\section*{Evaluation against independent observations}

Assessing the quality of a machine learning-based reanalysis poses unique challenges because the generated fields lie within the period covered by the training dataset. Large models have the capacity to memorise their training data, especially for observing systems with relatively small data volumes, such as radiosondes. To mitigate this risk, the training process was monitored using independent validation datasets to detect early signs of overfitting. Additionally, dropout strategies were implemented to prevent the model from relying too heavily on specific data types, forcing it to learn robust, generalised physical relationships rather than simply memorising localised inputs.

In this section, we evaluate AIFS-DOP using independent held-out datasets for the upper-air and the surface, and compare the results with established reanalysis products. The first evaluation considers the upper-air circulation. To this end, we use upper-air wind retrievals from the Multi-angle Imaging SpectroRadiometer (MISR)~\citep{mueller2017misr}, a multi-angle instrument aboard NASA's Terra satellite. These retrievals provide wind-vector estimates from cloud motion. Unlike many atmospheric motion vector (AMV) products, their height assignment is derived from stereoscopic geometry rather than from model fields. Since these observations were neither assimilated in ERA5 nor included in the AIFS-DOP training dataset, they provide a clean held-out validation target.

\begin{figure}[H]
  \centering
  \includegraphics[width=0.84\linewidth]{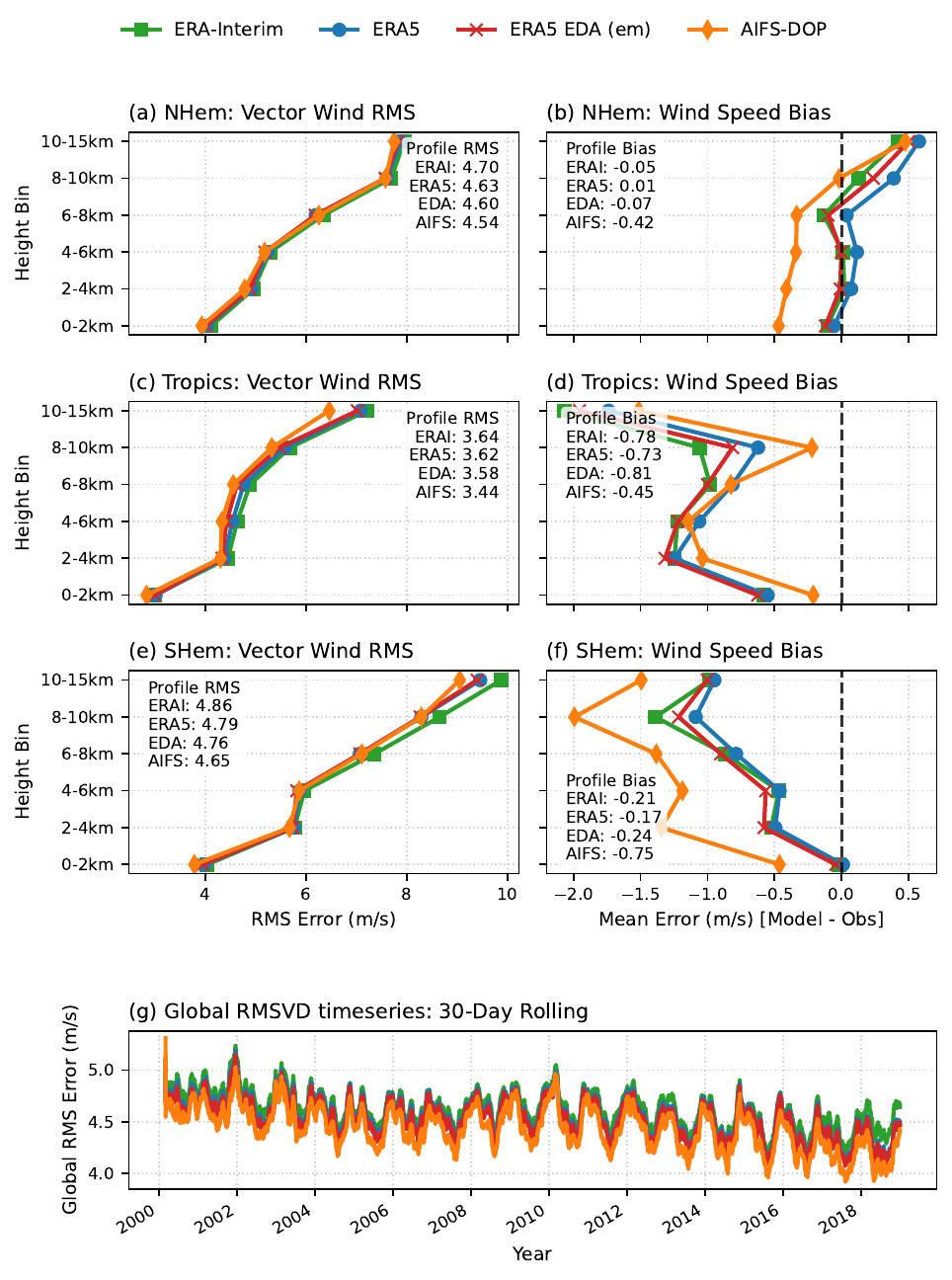}
  \caption{Validation of reanalysis wind fields against MISR AMV satellite observations (2000--2020). (a--f) Vertical profiles of vector wind RMS error (left) and wind speed bias (right) stratified by region: (a, b) Northern Hemisphere ($20^\circ\text{N}$ to $90^\circ\text{N}$), (c, d) Tropics ($20^\circ\text{S}$ to $20^\circ\text{N}$), and (e, f) Southern Hemisphere ($90^\circ\text{S}$ to $20^\circ\text{S}$) across six altitude bins up to 15 km for ERA5 (blue), ERA-Interim (green), the ERA5 EDA mean (red) and AIFS-DOP (orange). (g) Global 20-year time series of the 30-day rolling mean root mean square vector difference (RMSVD). All wind metrics are in ms$^{-1}$.}
  \label{fig:misr-evaluation}
\end{figure}

Figure~\ref{fig:misr-evaluation} shows the root mean square vector wind difference (RMSVD) and speed bias as a function of height for different latitude bands (a-f). The RMSVD values are found to be comparable to ERA5 and ERA-Interim while AIFS-DOP has a slow bias against MISR winds that is around 0.5 $ms^{-1}$  stronger compared to ERA5 in the extra-tropics. A time series of RMSVD is also shown in Figure~\ref{fig:misr-evaluation}g which confirms the temporal consistency of the AIFS-DOP reanalysis over 20 years with the quality remaining very similar to ERA5 throughout. Comparison against the ERA5 ensemble of data assimilation (EDA) mean is also shown.

Caution is needed before interpreting the similar (or lower) RMSVD errors in AIFS-DOP since the activity level of the analyses is different. ERA5 contains energy at small spatial scales that may be unconstrained by observations, as discussed in the previous section. This energy at small scales, while realistic, may penalise the ERA5 RMSVD values due to double penalty effects and may favour AIFS-DOP in this respect.

The second evaluation exploits the recent release of a new version of a comprehensive dataset of surface-land observations \cite{sfc-landv3}. It includes data sources previously unavailable to ECMWF reanalyses. Accordingly, we assess AIFS-DOP against these observations. For completeness, we show results for both (i) the entire dataset (many observations therein are included in the ECMWF archive), and (ii) only those observations that are independent. Independent observations are those falling within O96 grid cells with no matching correspondence in the ECMWF archive at that resolution, for each day of observation comparison (for the 24 hours of each day compared, plus the 12 hours that precede, to exclude any potential influence of preceding observations). This procedure ensures spatio-temporal exclusion, and the ECMWF archive is meant here as including surface (land and marine) as well as upper-air (balloon ascent) observations. The differences are then computed as the observations (considered at their original time and latitude, longitude location) minus the corresponding AIFS-DOP estimate (closest-in-time and nearest O96 gridpoint).

For comparison, we include previous-generation ECMWF reanalyses: ERA-15~\citep{era15}, ERA-40~\citep{era40}, ERA-Interim~\citep{erainterim}, and ERA5. We use these reanalyses at their native spatio-temporal resolution and apply bilinear spatial interpolation to the observation location (but no time interpolation). Note, we only consider here the 15th day each month for practical reasons. This approach is chosen rather than sub-sampling the dataset spatially, so as to capture the spatial variability of observations whose resolution may sometimes be far higher than O96.

\begin{figure}[H]
  \centering
  \includegraphics[width=0.84\linewidth]{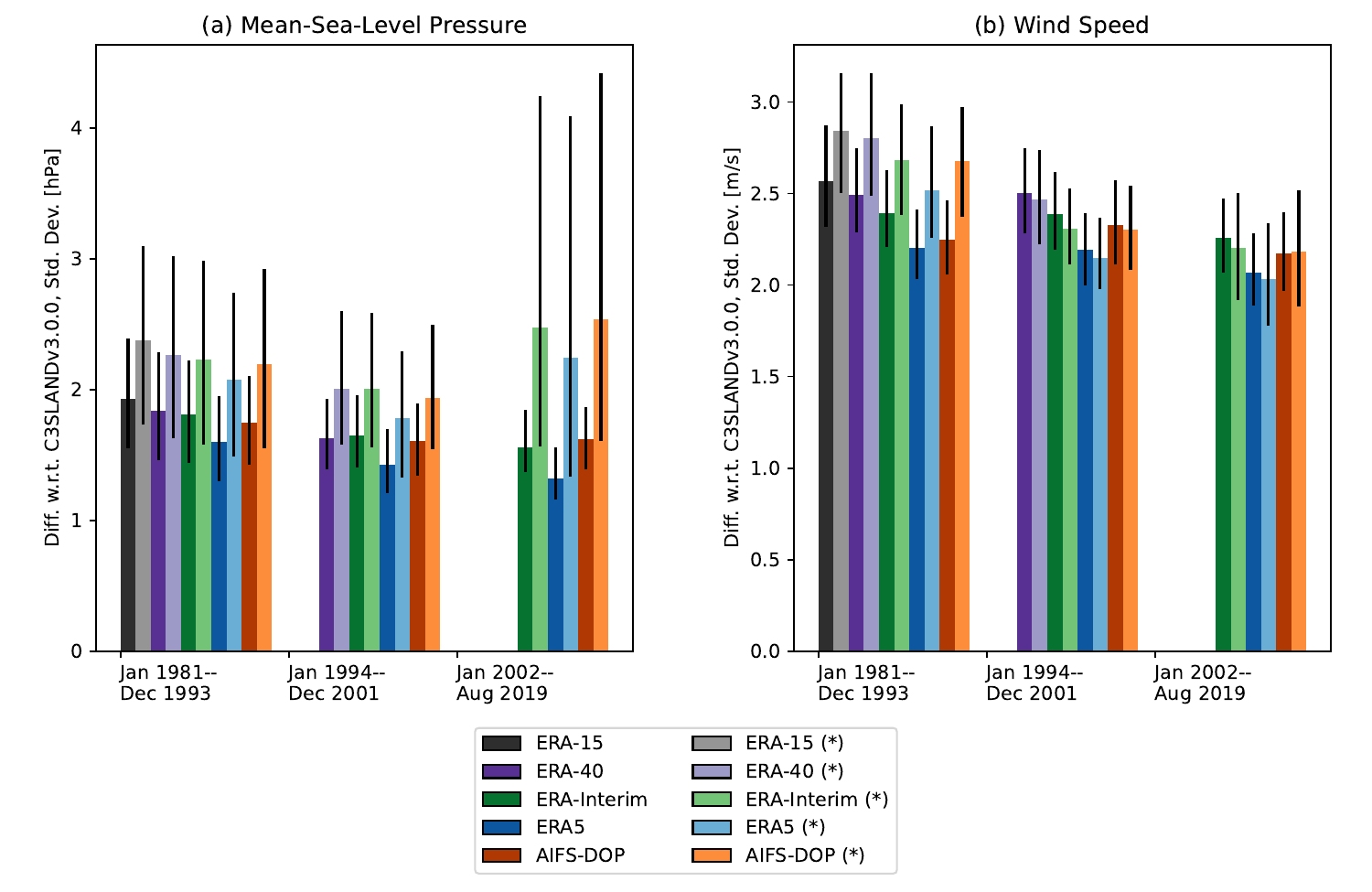}
  \caption{Assessment of AIFS-DOP (orange) against surface-land observations, showing the average of difference (interpreted as error) standard deviations within three selected time periods (see horizontal axis), for observables of the atmospheric circulation (a) mean-sea-level pressure and (b) wind speed. Previous generations of major ECMWF reanalyses are included in this assessment for comparison (see legend; ERA-15 in grey, ERA-40 in purple, ERA-Interim in green, and ERA5 in blue). The time periods reflect the availability dates of these prior reanalyses. Black vertical whiskers indicate the variability of the monthly difference standard deviations within each time period (10th and 90th percentiles). Dark colors show results including all the observations available, while lighter colors indicated by a star (*) in the legend show results for those observations that are only in the surface-land dataset and not included in any of the reanalysis datasets shown here.}
  \label{fig:sfcland-evaluation}
\end{figure}

Figure~\ref{fig:sfcland-evaluation} shows the standard deviations of differences (interpreted here as errors), obtained with this method. The whiskers indicate non-negligible variability and suggest to exercise caution when interpreting the results. However, the performance ranking of the four successive generations of reanalyses is consistent with well-established findings, with errors reducing from ERA-15, to ERA-40, then ERA-Interim, and finally ERA5. Given this remark, the hypothesis that the evaluation method and data sample are representative of established results cannot be discarded, and is assumed to hold hereafter.

The variations observed between the time periods generally reflect the increasing performance of reanalyses over time as the observing system improves. The locations of the observations selected for assessment also have an influence, particularly if one considers only the independent observations (lighter colors), which, by selection, cover regions where none of the reanalyses have corresponding in-situ observation input. The performance of AIFS-DOP is generally on par with ERA-Interim or better. This would suggest that the machine learning approach presented in this paper places the resulting products' performance between ECMWF 4th- and 5th-generation reanalyses, at least for the surface parameters investigated here.

\section*{Discussion and conclusions}

Traditional data assimilation systems use a physical model and covariance statistics to produce dense, physically coherent fields from sparse observations. Here we have presented results which suggest that the representations of the physical Earth system learned during training perform a similar role in observation-driven machine learning models.

While traditional reanalyses typically take several years to produce, the reanalysis presented here was computed during the course of a single working day. The training for this model took four days running on 16 H100 GPUs, and the reanalysis inference was run in eight parallel streams. While this model was relatively low resolution, for a model with N320 input/output resolution, we expect the training costs to be approximately 50\% higher than at O96 and inference could still be completed within a day using additional parallel streams. This step-change in production time potentially opens a new paradigm of iterative refinement that would previously have been impossible. Ensembles of reanalyses and high resolution nesting for regional reanalysis become options that can be considered.

Unlike hybrid approaches (such as diffusion based data assimilation techniques) which train on ERA5 data, the reanalysis in this paper is not an emulation of ERA5 but instead a completely independent climate reconstruction.

It is important to note that other observation-only records (e.g., HadCRUT5 \citep{Moriceetal2021}, COBE-SST \citep{Ishiietal2005}, GISTEMP \citep{Lenssenetal2019}) also deliver vital historical baselines but are usually constrained to individual fields, primarily near-surface temperature. These datasets are based on geostatistical interpolation methods, such as Kriging or Optimal Interpolation, to reconstruct complete fields from sparse observing networks. Consequently, they do not attempt to synthesise the comprehensive, multi-component representation of the Earth system that the AIFS-DOP framework aims to establish.

By necessity, this paper has covered a broad range of diagnostics to demonstrate that this machine-learning-generated reanalysis performs encouragingly well across a wide spectrum of use cases. Trust in the ERA5 dataset has been built up over years through countless research papers; a single paper could never exhaustively cover all these aspects. The aim of this paper was to demonstrate the promise of the \textit{method} rather than to introduce this \textit{dataset} as a new reanalysis product in its own right. We believe that we have shown that machine learning from observations alone is a viable approach for reanalysis generation that merits further investigation.

The ability of the model to reconstruct the mean state and the dominant modes of variability is reassuring. However, the presence of features such as ENSO teleconnection patterns (Figure~\ref{fig:extended-data-enso-teleconnections}) does not necessarily mean that the model has learned a representation of these processes; it may simply indicate that the observations are sufficiently dense to constrain these features at initialisation time.

The results in this prototype, while highly encouraging, also reveal that there is still more work to be done to reach the full fidelity of state-of-the-art reanalysis products. A number of issues still need to be resolved such as the representation of the meridional circulation at mid-levels in the tropics and the predictions of relative humidity over dry polar regions. Further evaluation will likely reveal other deficiencies which will need to be addressed through further changes in the data quality control, training procedure or model architecture.

A known consequence of training with a mean squared error loss is that the model is encouraged to predict the conditional mean of the atmospheric state. Where the observations do not fully constrain small scale or unpredictable rapidly evolving features, this can lead to smoother fields and reduced spatial variance. This should be regarded as a limitation for applications that require realistic instantaneous mesoscale structures. However, many climate use cases of reanalysis (such as those shown here) are concerned primarily with the mean state, large-scale variability, and slowly varying anomalies rather than with the exact phase of unpredictable small scale features. For such uses, estimating the predictable mean state may be an appropriate and even desirable objective. Training with a probabilistic loss would be expected to produce realistic levels of activity at all scales. Possible approaches include diffusion-based training \citep{Price2025GenCast} or directly optimising a loss based on a proper scoring rule \citep{lang2026aifs}. Crucially, probabilistic training will make it possible to provide uncertainty estimates. Because (once trained) the model is computationally cheap to run, potentially a large number of ensemble members can be generated.

This study did not include many core observation categories. Global Navigation Satellite System Radio Occultation (GNSS-RO), hyperspectral infrared, microwave imager, scatterometer and other platforms known to improve skill in traditional numerical weather prediction (NWP) systems will be added in future upgrades. In addition, the DOP architecture makes it easy to make use of observation types that are more challenging to include in physics-based systems. Furthermore, since the DOP approach naturally extends to fully coupled modelling \citep{boucher2025coupled}, observations from the sub-surface ocean, land surface and atmospheric composition can be included. Improvements can also be expected from increasing the spatial and temporal resolution of the model. 

Observation-driven frameworks offer a practical way to extend the reach of short-term observation campaigns across much longer periods. When a specific variable is measured during a brief field campaign, those targeted observations overlap with continuous data from the rest of the observing system, such as satellites and permanent ground networks. During this concurrent window, the DOP model can learn the underlying correlations between the campaign variable and the broader network. The trained model can use the historical record of the permanent observing system to reconstruct the campaign variable across the entire time period. Work is ongoing to explore the potential role of campaign data in observation-driven models.

Given the important role that reanalysis datasets play in the detection of relatively small climatic trend signals from often highly variable fields, further work is needed to validate the robustness of machine learning models such as AIFS-DOP to the non-stationary observing system. For example, experiments examining the robustness of the predicted fields to time varying instrument biases would be important if future machine-learning-generated reanalyses are to be considered suitable for climate-grade decision making.

In our opinion, fully data-driven machine learning techniques have a role to play in the field of reanalysis production. They offer state estimates which are independent from physics-based numerical models and open up new possibilities that could ultimately enhance our ability to study and understand atmospheric processes.

\begin{table}[htbp!]
\caption{Description of curated observation dataset}
\label{tab:dataset_description}
\centering
\begin{tabular}{|p{2.5cm}|p{2.5cm}|p{2cm}|p{2.5cm}|p{4.3cm}|}
\hline
\textbf{Category} & \textbf{Instruments} & \textbf{Period} & \textbf{Variables} & \textbf{Details} \\
\hline
Infrared Sounder & HIRS & 1980--2022 & Brightness Temperatures & EUMETSAT Fundamental Data Record \citep{eumetsat_hirs_0961} \\
\hline
Microwave Sounders & MSU \newline SSM/T-2 \newline AMSU-A \newline AMSU-B \newline MHS \newline ATMS & 1980--2005 \newline 1994--2005 \newline 1998--2022 \newline 1998--2014 \newline 2005--2022 \newline 2012--2022 & Brightness Temperatures & MSU taken from NOAA Climate Data Record \citep{zou_msu_2013}, SSM/T2 taken from EUMETSAT Fundamental Data Record \citep{eumetsat_ssmt2_0304} \\
\hline
Surface Observations & SYNOP, Buoys, Ships & 1980--2022 & 2t, 2d, msl, 10u, 10v, sst & Existing ECMWF data archive \\
\hline
Upper-air Observations & Radiosonde, Aircraft, AMV & 1980--2022 & t, u, v, z, q on pressure levels & Existing ECMWF data archive\\
\hline
Geostationary Satellite & GridSat & 1980--2022 & Brightness Temperatures & NOAA Climate Data Record \citep{knapp_gridsat_2011} \\
\hline
\end{tabular}
\end{table}

\begin{table}[ht]
\centering
\caption{Instrument name definitions.}
\label{tab:instrument_names}
\begin{tabular}{ll}
\toprule 
Acronym & Full name \\
\midrule
MSU & Microwave Sounding Unit \\
AMSU-A & Advanced Microwave Sounding Unit-A \\
ATMS   & Advanced Technology Microwave Sounder \\
SSM/T-2 & Special Sensor Microwave Humidity-2 \\
AMSU-B & Advanced Microwave Sounding Unit-B \\
MHS    & Microwave Humidity Sounder \\ 
HIRS & High Resolution Infrared Radiation Sounder \\
AMV & Atmospheric Motion Vectors \\
\bottomrule
\end{tabular}
\end{table}

\begin{table}[htbp]
    \centering
    \caption{Summary of the predicted surface and upper-air variables.}
    \label{tab:prediction_variables}
    \begin{tabular}{@{}llp{5.5cm}@{}}
        \toprule
        \textbf{Domain} & \textbf{Variables} & \textbf{Pressure Levels (hPa)} \\
        \midrule
        \textbf{Upper Air} 
        & Temperature ($t$) & 100, 200, 250, 300, 400, \\
        & U-component of wind ($u$) & 500, 700, 850, 925 \\
        & V-component of wind ($v$) & \\
        & Specific humidity ($q$) & \\
        & Geopotential ($z$) & \\
        \midrule
        \textbf{Surface} 
        & Mean sea level pressure (msl) & Single level \\
        & Surface pressure (sp) & \\
        & 10-metre U wind component (10u) & \\
        & 10-metre V wind component (10v) & \\
        & 2-metre temperature (2t) & \\
        & 2-metre dewpoint temperature (2d) & \\
        & Sea surface temperature (sst) & \\
        \bottomrule
    \end{tabular}
\end{table}

\clearpage

\section*{Methods}

\subsection*{Effective Coriolis parameter diagnostic}

The effective Coriolis parameter diagnostic used in Figure~\ref{fig:coriolis-parameter} is derived as follows. Starting with geostrophic balance defined as

$$u = -\frac{1}{f} \frac{\partial \Phi}{\partial y}, \quad v = \frac{1}{f} \frac{\partial \Phi}{\partial x}$$

which defines the relationship between the zonal wind component, \textit{u}, the meridional wind component, \textit{v}, and the geopotential, $\Phi$. Solving for the Coriolis parameter, \textit{f},

$$f \approx \frac{v \frac{\partial \Phi}{\partial x} - u \frac{\partial \Phi}{\partial y}}{u^2 + v^2}$$

$$\bar{f} = \frac{\sum_{t} \left( v_t \frac{\partial \Phi_t}{\partial x} - u_t \frac{\partial \Phi_t}{\partial y} \right)}{\sum_{t} \left( u_t^2 + v_t^2 \right)}$$

\section*{Acknowledgements}

Florence Rabier is thanked for her support of the DOP project from the outset. We thank NOAA and EUMETSAT for their provision of high quality satellite datasets and in particular Viju John and Roope Tervo for their advice and assistance. The authors gratefully acknowledge the Gauss Centre for Supercomputing e.V. (www.gauss-centre.eu) for funding this project by providing computing time on the GCS Supercomputer JUPITER at Jülich Supercomputing Centre (JSC). Ewan Pinnington's contribution is funded by the CERISE project (grant agreement No101082139). CERISE is funded by the European Union. Views and opinions expressed are however those of the author(s) only and do not necessarily reflect those of the European Union or the Commission. Neither the European Union nor the granting authority can be held responsible for them.

\bibliographystyle{unsrtnat}
\bibliography{references}

\end{document}